%% file: DraftToArchive.tex
\documentclass[twocolumn,showkeys]{revtex4}
\usepackage[pdftex]{graphicx}
\usepackage{amssymb,amsfonts,amsmath}
\usepackage{pdfpages}
\begin{document}

%%%%%%%%%%%%%%%%%%%%%%%%%%%%%%
\begin{abstract}
The process of quantum measurement has been a long standing source of debate. A measurement is postulated to collapse a wavefunction onto one of the states of a predetermined set - the measurement basis. This basis origin is not specified within quantum mechanics. According to the theory of decohernce \cite{ZurekRMP}, a measurement basis is singled out by the nature of coupling of a quantum system to its environment. Here we show how a measurement basis emerges in the evolution of the electronic spin of a single trapped atomic ion due to spontaneous photon scattering. Using quantum process tomography we visualize the projection of all spin directions, onto this basis, as a photon is scattered. These basis spin states are found to be aligned with the scattered photon propagation direction. In accordance with decohernce theory, they are subjected to a minimal increase in entropy due to the photon scattering, while, orthogonal states become fully mixed and their entropy is maximally increased. Moreover, we show that detection of the scattered photon polarization measures the spin state of the ion, in the emerging basis, with high fidelity. Lastly, we show that while photon scattering entangles all superpositions of pointer states with the scattered photon polarization, the measurement-basis states themselves remain classically correlated with it. Our findings show that photon scattering by atomic spin superpositions fulfils all the requirements from a quantum measurement process.
\end{abstract}
\date{\today}
\title{Emergence of a measurement basis in atom-photon scattering}

\author{Yinnon Glickman}
\email[E-mail: ]{yinnon.glickman@weizmann.ac.il}
\author{Shlomi Kotler}
\author{Nitzan Akerman}
\author{Roee Ozeri}
\affiliation{Department of Physics of Complex Systems, Weizmann Institute of Science, Rehovot 76100, Israel}

\maketitle

%Quantum measurement and the origins of classical behavior
One of the controversial postulates of quantum mechanics is the measurement postulate. Following a measurement, a quantum superposition is postulated to collapse on a basis that is determined by the measurement apparatus. One of the fundamental questions posed by wavefunction collapse pertains to the origin of the measurement basis. What is it in the physical world that ``chooses'' the specific basis onto which the system collapses?

According to the theory of decoherence, initially introduced by Zurek \cite{Zurek1981,ZurekRMP}, the key lies outside the measured system. A quantum system is immersed in and interacts with an environment that contains the measurement device and possibly additional degrees of freedom. Rather than being arbitrarily determined by an obscure observer, a measurement (pointer) basis is singled out (einselected) by the coupling Hamiltonian between the system and the environment degrees of freedom. During the process of a measurement the system dynamically evolves towards one of these states. The repeatability postulate of quantum mechanics asserts that repeated measurements in the same basis render the same result; i.e. if the system collapsed to the state $|s_k\rangle$ then additional measurements in the same basis will always point to $|s_k\rangle$; system states that change due to the interaction with the environment cannot qualify as pointer-basis states. Only system states that remain unchanged can be considered as possible candidates for a measurement basis. Moreover, following the measurement process, these basis states remain disentangled from the environment while any other state will end up entangled with it. Formally, a way to find pointer states is by searching for states that do not gain entropy following coupling to the environment; the so-called predictability sieve \cite{ZurekRMP}.

%Relation to quantum-to-classical transition; photon scattering as a "general measurement"
Decoherence theory was not only successful in explaining the origin of the emergent measurement basis, but it also shed light on the quantum-to-classical transition. While microscopic bodies, such as electronic spins, can coexist in a superposition of states \cite{SternGerlach}, macroscopic bodies are always found in a single classical state. According to decoherence theory macroscopic bodies are continuously measured by their environment, and as a result, constantly pinned to a well determined classical state. The position of a macroscopic body that is exposed to light, for example, is continuously measured by photons that scatter from its surface. In other words, we can constantly ``see'' where it is \cite{ReidelZurek2010}.

Since atomic systems can be well isolated from their environment and coherently controlled with good fidelity, they are a good experimental platform for the study of such fundamental quantum phenomena. In a typical experiment, a bipartite atomic superposition is controllably coupled to its environment and monitored in order to investigate different facets of the quantum-to-classical transition such as decoerence, measurement, and system-environment entanglement. Experiments have investigated the decoherence of an atomic superposition due to coupling to engineered reservoirs using trapped atomic ions \cite{Myatt00}. The interaction between atoms and a microwave cavity was used to observe the progressive decoherence of the measurement apparatus in quantum measurement \cite{Haroche96}.

A particularly interesting example is the coupling between atomic systems and the electromagnetic vacuum via spontaneous photon scattering. The relation between the information carried by the scattered photon and the decoherence rate was established in numerous studies. The effect of light scattering on the coherence of atomic interferometers, where scattered photons expose the path an atom has taken, was investigated \cite{ozeri2005, Uys2010, Akerman2011}. Photon scattering by trapped atomic ions, where the direction and magnitude of the internal angular momentum of an atom become correlated with a scattered photon, was shown to result in spin decoherence [10, 11, 12]. Photon scattering is also a widespread tool to measure the state of atomic superpositions. State-selective florescence using resonant laser light was used to measure the internal electronic state of atoms with a very small error probability \cite{Hume07, Lucas08, Anna11}. Finally, the entanglement between a single atom and a spontaneously scattered photon was recently observed \cite{Monroe04, Weinfurter06}.

In this work, we demonstrate that decoherence, quantum measurement and atom-photon entanglement are intertwined in the photon scattering process. Here, we use the spin of a single atomic ion as a quantum system that is coupled to its environment by photon scattering. Quantum process tomography (QPT) \cite{IkeAndMike} is applied to monitor the spin state evolution under photon scattering. QPT is a protocol for obtaining the quantum operation elements that act on a certain quantum system. It is based on probing the process with a known set of initial quantum states and analyzing, using quantum state tomography, the corresponding post-process final states. QPT was applied for characterization of quantum gates in trapped ions platform \cite{BlattQPTGate}, teleportation channel with atoms \cite{BlattTeleportationQPT} and Josephson-phase qubit memory \cite{KatzQPT} among many other experiments. The application of QPT allowed us to show how, following photon scattering, the Bloch sphere, representing  all possible pure spin directions, collapses onto an elongated spheroid that is aligned with a pointer-state basis direction. We further show that the pointer states are those which gain the least amount of entropy in the scattering process. By performing a measurement of the scattered photon polarization one can measure the spin state of the ion, in the pointer-state basis, with good fidelity. Lastly, we show that, while all initial spin states become correlated with the scattered photon polarization, pointer states are classically correlated with it whereas their superpositions are entangled with it. We thus show that the process of photon scattering by an atomic superposition satisfies all the requirements from a quantum measurement process, as determined by decoherence theory.

\section{Photon scattering process}
To study the electronic spin evolution in the process of photon
scattering we resonantly scatter photons on the $5\
s^2S_{1/2}\rightarrow\ 5\ p^2P_{1/2}$ transition in a single trapped
$^{88}$Sr$^{+}$ ion, shown diagrammatically in the inset of Fig. 1. On
this transition, both ground and excited states are spin 1/2
manifolds. It is therefore convenient to think of the coupling between these manifolds in terms of spin 1/2 (Pauli) operators. In events where a single photon was emitted into a direction, $\hat{k}$, the
spin degrees of freedom in the ground and excited states are coupled
via a single application of the spin operator $i(\vec{\sigma} \cdot
\vec{E})^\dagger$. Here $\vec{\sigma}$ is a vector of the Pauli spin 1/2
operators and $\vec{E}$ is the polarization vector of the emitted photon.
In the case of absorption the coupling operator is $-i\vec{\sigma} \cdot
\vec{E}$. Derivation of these operators is given in the supplementary material.

Consider the ion in the electronic excited state with a spin
pointing in a general direction and no photons in mode $\vec{k}$,
$|\psi_i\rangle = |\phi\rangle_{spin}\otimes|0\rangle_{\vec{k}}$.
It can be shown that a single-photon
 spontaneous emission acts on $|\psi_i\rangle$ with the operator
\begin{equation}
\label{Hint} U_{int} = (\vec{\sigma}\cdot\vec{E}_1)^\dagger\otimes
a^\dagger_{\vec{k},\vec{E}_1} + (\vec{\sigma}\cdot\vec{E}_2)^\dagger\otimes
a^\dagger_{\vec{k},\vec{E}_2},
\end{equation}
where $\vec{E}_1$ and $\vec{E}_2$ are a basis for the emitted photon
polarization and $a^\dagger_{\vec{k},\vec{E}_i}$ are the respective
photon creation operators. Since the photon electric field lies in the plane normal to $\vec{k}$ it is spanned by two polarization components. Therefore, although a general polarization state is spanned by the three Stocks parameters, this discussion is restricted to two polarization basis states only.

If $\vec{E}_1$ and $\vec{E}_2$ are real
vectors, i.e. represent a linear polarization basis, then the effect
of photon emission is that of unitary rotations that are correlated
with orthogonal photon polarization states. These rotations can be
subsequently reversed \cite{Akerman2011}. A choice of a complex
polarization basis, e.g. orthogonal circular polarizations, leads to
a pair of non-unitary, and therefore irreversible, spin operators.
The reduced spin density matrix following photon emission is, of
course, independent of the photon polarization basis choice.

It is, however, instructive to study the reduced spin evolution using a
circular photon polarization basis. In this case, the two spin operators that correspond to the scattered photon polarization components in
Eq.\ref{Hint}, $\vec{\sigma}\cdot\vec{E}_\pm ^*= \sigma_\mp$, are the
spin ladder operators in the $\vec{k}$ direction. This means that
spin states that initially point along the emitted photon
propagation direction can emit circularly polarized photons but only with the helicity parallel
their spin. Following emission of a circularly polarized photon, the spin direction reverses but
remains aligned with the photon $\vec{k}$ vector. This is a simple
manifestation of angular momentum conservation in the photon
emission process. Using the spin $\vec{k}$ direction and circular photon polarization as a basis, the combined spin-photon state following scattering can be written as,
\begin{equation}
\label{Final state} U_{int}(\alpha|\hat{k}\rangle+\beta|-\hat{k}\rangle)\otimes|0\rangle_{\vec{k}} = \alpha|-\hat{k}\rangle\otimes|E_+\rangle_{\vec{k}}+\beta|\hat{k}\rangle\otimes|E_-\rangle_{\vec{k}}.
\end{equation}
Here $|\pm\hat{k}\rangle$ represent states of a spin pointing in the $\pm \hat{k}$ direction and $|E_{\pm}\rangle_{\vec{k}}$ represent states of a single, right or left circularly polarized, photon with wave vector $\vec{k}$.

Spin states pointing in the $\hat{k}$ direction therefore constitute a good pointer-state basis. They are classically correlated with the emitted photon polarization, whereas any superposition of such states becomes entangled with it. Starting from a superposition of spin pointer states and tracing over the scattered photon polarization the spin state will evolve into a statistical mixture. When starting in this spin direction, however, photon scattering, theoretically, does not induce entropy increase \footnote{Notice that, unlike the more common
treatment of a pointer-state basis \cite{ZurekRMP}, the $|\pm\vec{k}\rangle$
states are not eigenstates of, and therefore are not invariant
under, $U_{int}$. They are, however, connected to $U_{int}|\pm\vec{k}\rangle$ by a unitary transformation that does not increase entropy. This observation is also valid for the case of absorption of a linearly polarized photon and the subsequent emission of a photon in a general direction.}.

To evaluate spin evolution in the full scattering process, one has
to account for photon absorption as well. Here, spin evolution
depends on the polarization of the excitation laser and its
direction relative to that of the scattered photon. In our
experiment the ion  is excited with a linearly polarized laser beam (in
the $\hat{z}$ direction). Photons are detected in a direction
perpendicular to that of the beam polarization ($\hat{x}$ direction; See
Fig. 1). Absorption of a photon is accompanied by an
application of $\vec{\sigma}\cdot\vec{E}_z = \sigma_z$. The
spin-photon combined state evolves under,
\begin{eqnarray}
\label{X} U_{int} = \vec{\sigma}_{(+,x)}\vec{\sigma}_z\otimes
a^\dagger_{\vec{k},\vec{E}_+} +
\vec{\sigma}_{(-,x)}\vec{\sigma}_z\otimes
a^\dagger_{\vec{k},\vec{E}_-} \\ \nonumber = |-\hat{x}\rangle\langle
-\hat{x}|\otimes a^\dagger_{\vec{k},\vec{E}_+}+|\hat{x}\rangle\langle \hat{x}|\otimes
a^\dagger_{\vec{k},\vec{E}_-},
\end{eqnarray}
where $\vec{\sigma}_{(\pm,x)} = \sigma_y \mp i\sigma_z$ are the spin raising and lowering
operators in the $\hat{x}$ direction. The spin of the ion is therefore
projected along the emitted photon direction every time a photon
with a circular polarization is detected. The post-scattering spin
state, calculated by tracing over the scattered photon
degrees of freedom, is
\begin{equation}
\label{tracingOver}
 \rho_{spin}= |\alpha|^2| \hat{x} \rangle \langle \hat{x}
| + |\beta|^2| -\hat{x} \rangle \langle -\hat{x} |.
\end{equation}
Here $\alpha$ and $\beta$, are the initial spin amplitudes in the
$\hat{x}$ and $-\hat{x}$ directions respectively. Equation \ref{tracingOver}
represents, in general, an incoherent mixture of states, into which
the initial spin states evolve. It suggests that the spin state will
decohere unless it is initialized in the $| \hat{x} \rangle$ or $|
-\hat{x} \rangle$ states.

\begin{center}
\begin{figure}
\includegraphics[scale=0.6]{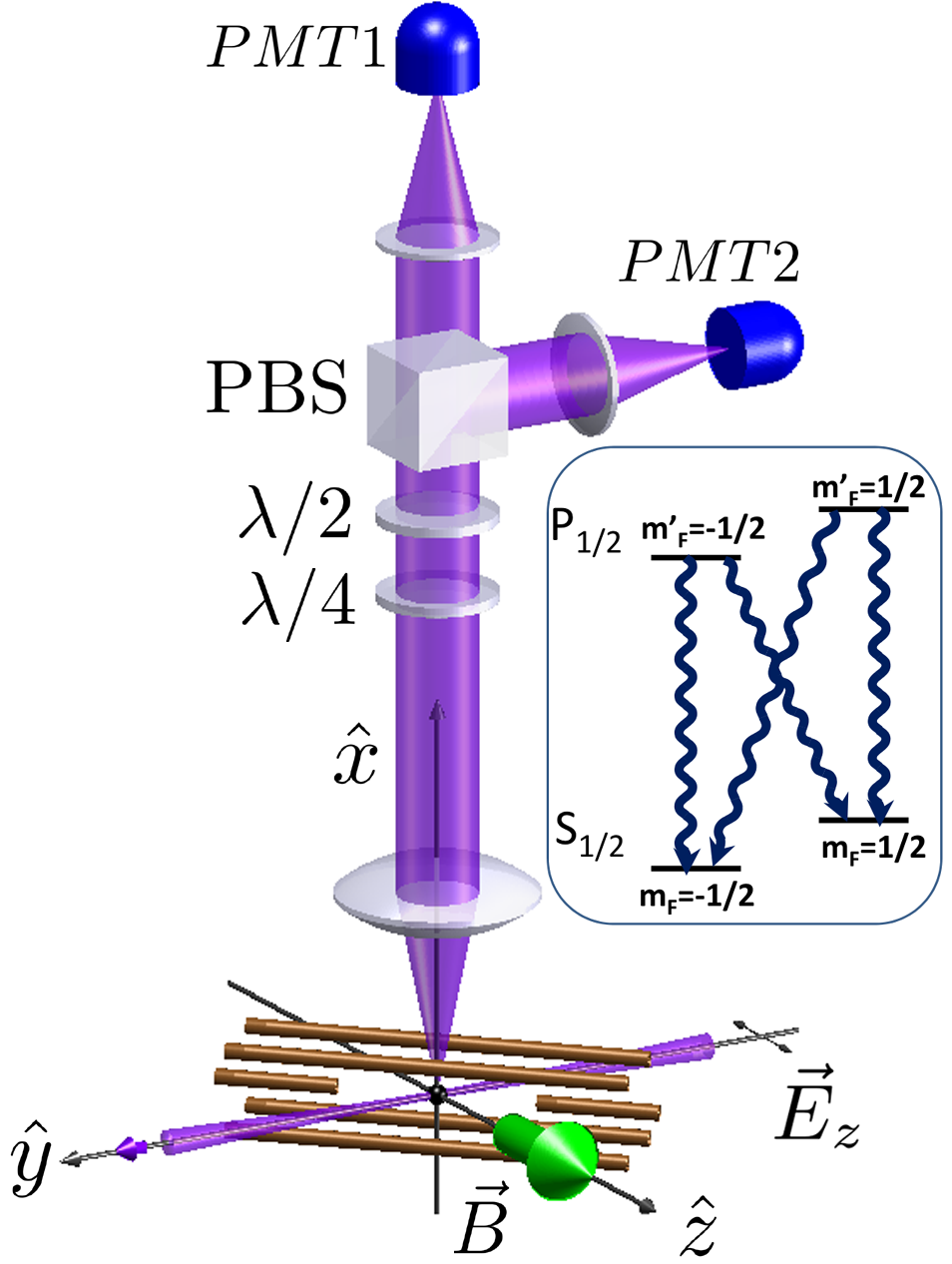}
\caption{Schematic drawing of the experimental setup. A single $^{88}$Sr$^+$ ion is trapped in a linear Paul trap \cite{TrapPaper11}. A magnetic field $\vec{B}||\hat{z}$ Removes the degeneracy between the two spin states of the valence electron in the $5\ s^2S_{1/2}$ ground level by $\hbar\omega_0$ where $\omega_0/2\pi = 3.5$ MHz. The inset shows a schematic level structure of the $5 s^2S_{1/2}$ and the
$5\ p^2P_{1/2}$ electronic levels. The 422 nm transition between these two levels is excited using a weak resonant beam, propagating along the $\hat{y}$ direction and linearly polarized along $\hat{z}$ ($\vec{E}_z$). Scattered photons are collected along the $\hat{x}$ direction and their polarization is analyzed using half- and quarter- wave retardation plates and a polarizing beam splitter (PBS). The two ports of the PBS are directed towards two photo multiplier tube detectors. With this setup the emitted photon polarization can be measured in any orthogonal basis.} \label{setup}
\end{figure}
\end{center}

\section{Experimental setup}
A schematic illustration of our experimental setup is shown in Fig. 1. A single $^{88}$Sr$^{+}$ ion was trapped and laser-cooled in a linear Paul trap. A weak magnetic field, applied in the $\hat{z}$ direction, removed the degeneracy between the two spin states of the valence electron in the $5\ s^2S_{1/2}$ ground level by $\hbar\omega_0$ where $\omega_0/2\pi = 3.5$ MHz. The ion spin was polarized by means of optical pumping to the $|\uparrow\rangle = |\hat{z}\rangle$ state. Spin readout was performed using shelving of the $|\uparrow\rangle$ state to the metastable $4\ d^2D_{5/2}$ level, followed by state-selective fluorescence. Spin rotations, in the $5\ s^2S_{1/2}$ manifold, were performed using resonant radio-frequency (RF) pulses. For more details see \cite{Anna11, TrapPaper11}. After optical pumping the spin was oriented in any desired direction by applying the appropriate RF pulse. To ensure that the spin direction is well defined in the lab frame of reference we reset the phase of our RF oscillator at the beginning of each repetition of the experiment. Subsequent to this initialization procedure the spin performed Larmor precession around the $\hat{z}$ direction at $\omega_0$.

% describe the ion photon interaction
Following spin state preparation a, weak, 422 nm, laser beam resonant with the $5\ s^2S_{1/2}\rightarrow 5\ s^2P_{1/2}$ transition (see the inset in Fig. 1), and polarized in the $\hat{z}$ direction, was turned on for a fixed time. The ion scattered a photon from this beam with a probability between 0.05-0.1. Since our Zeeman splitting was small compared with the transition linewidth (21 MHz) the two spin states scattered photons with approximately the same probability. Photons were detected from a direction perpendicular to the excitation laser direction ($\hat{x}$ direction). A 0.31 N.A. objective lens \cite{Alt} collected roughly 1\% of the scattered photons.  A polarizing beam splitter (PBS) directed vertically and horizontally linearly-polarized photons to two separate photo multiplier tubes (PMT's). Half- and quarter-wave retardation plates preceding the polarizer allowed for photon polarization measurement in any desired basis.  The overall detection efficiency of scattered photons was measured to be roughly 1/400. In all the experiments reported here, we post-selected only those repetitions of the experiment where the PMTs indicated that a single photon was recorded. In 85-90\% of these events a single photon was indeed scattered, whereas in the remaining 10\%-15\% of events detection was generated either by a dark count of the PMT or a single detected photon out of a scattered pair.

To investigate spin dynamics due to photon scattering, it is important to know the spin direction within the Larmor precession cycle at the moment the photon was scattered. To this end, we recorded the phase of our local oscillator, tuned to the Larmor precession frequency $\omega_0$, at the time the scattered photon was recorded by the PMT \cite{Akerman2011}.

\begin{figure}[!htb]
\begin{center}
\begin{tabular}{ccc}
\includegraphics[scale=0.13]{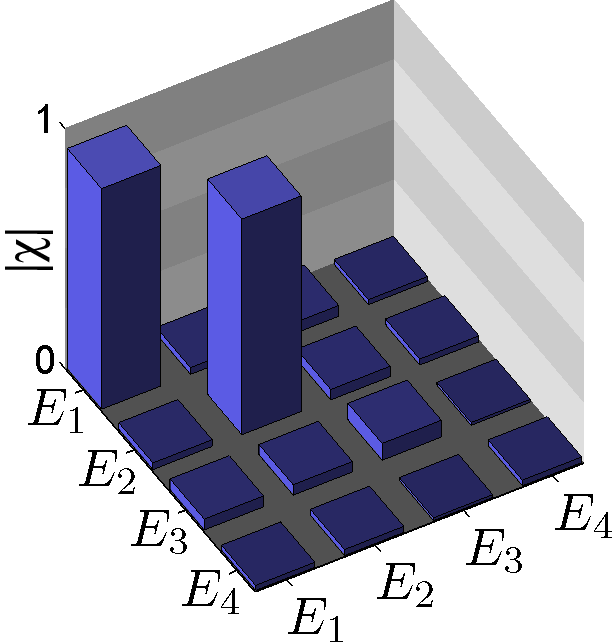} &
\includegraphics[scale=0.13]{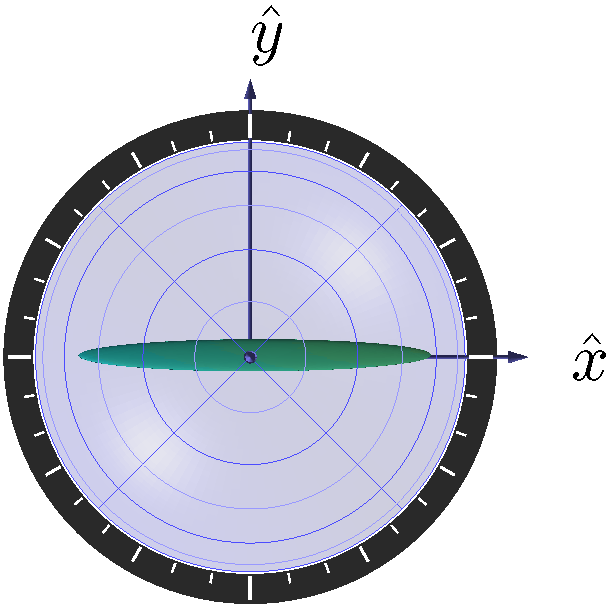} &
 \includegraphics[scale=0.12]{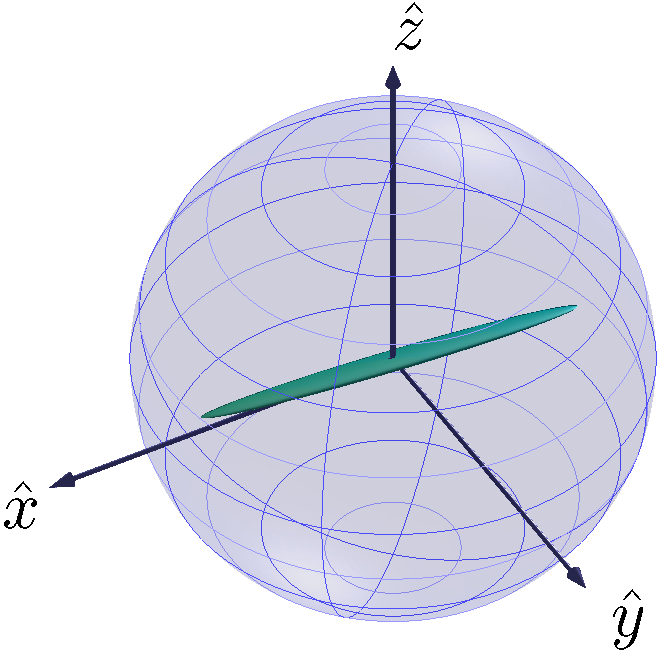}\\
(a) & (b) & (c)\\
\multicolumn{3}{c}{(d)\includegraphics[scale =
0.22]{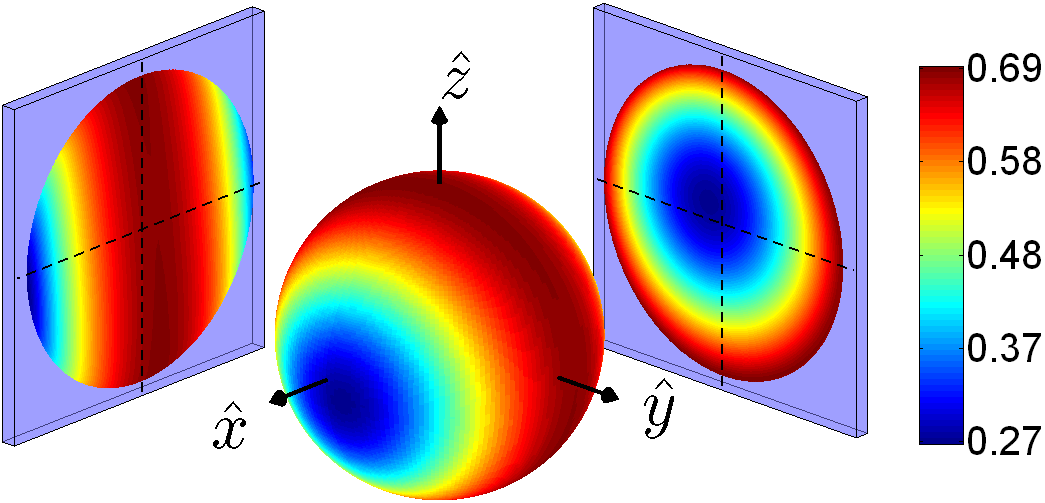}}
\end{tabular}
\caption{Collapse of the Bloch sphere.(a) The absolute value of the elements of the reconstructed process matrix. The basis used here is $E_1=|\hat{x}\rangle \langle\hat{x}|=\left(I+\sigma_x\right)/2$, $E_2=|-\hat{x}\rangle \langle-\hat{x}|=\left(I-\sigma_x\right)/2$ (projections on the $|\pm \hat{x}\rangle$ states), $E_3=-i\sigma_y$ and $E_4=\sigma_z$. As seen, the process is mainly composed of equal contributions of projections on the $\hat{x}$ direction. Two view points; (b) along the $z$ axis  and (c) at 45$^\circ$ to the $\hat{z}$ axis of the surface on which the Bloch sphere of initial spin states collapses, following photon scattering. The resulting elongated ellipsoid, clearly marks the emergence of a spin measurement basis. (d) The von Nuemann entropy of post-scattering spin states. This is also the increase in entropy due to photon scattering. The states on the tip of the emergent basis (pointer states) acquire the minimal amount of entropy, in accordance with the predictability sieve criteria in decoherence theory. Equal superpositions of pointer states, on the other hand, acquire almost the maximal possible entropy demonstrating that, without any knowledge on the scattered photon, the process of photon scattering is in general irreversible.}\label{BlochSphereCollapse}
\end{center}
\end{figure}

\section{Collapse of the Bloch sphere}
% elaborate on the quantum process tomography
Following a scattering event the spin state is expected to project on either
the $|\hat{x}\rangle$ or the $|-\hat{x}\rangle$ states (\ref{X}). Therefore,
all initial pure spin states, that are not aligned with the $\pm \hat{x}$
directions, are expected to decohere into statistical mixtures of
these states. In the Bloch sphere geometric representation this
operation maps the surface of the sphere onto the $\hat{x}$ axis. We
monitored this collapse of the Bloch sphere using QPT.

In single-qubit QPT, state tomography is performed on the post-scattering states of four, linearly independent, initial spin states, requiring a total of 12 different measurement types. Experiments were repeated until 900 successful repetitions, i.e. ones in which a photon was detected, were recorded for each measurement type. Typically, process tomography is performed in a frame of reference that is rotating along with the spin. However, spin evolution due to photon scattering is determined by directions in the lab frame of reference. To faithfully perform state tomography in the lab frame we perform state tomography stroboscopically with the spin Larmor precession. To this end, rather than analyzing all recorded events, we limited our analysis to events that occurred within a $2\pi/32$ radian phase interval. The direction of the spin at the moment of scattering in these events was spread over an angular span which equals this phase interval around the $\hat{x}$ direction. Choosing other phase intervals of similar width yield identical results; further details are given in the supplementary material.
\begin{figure}
\begin{center}
\begin{tabular}{cc}
(a)\includegraphics[scale = 0.15]{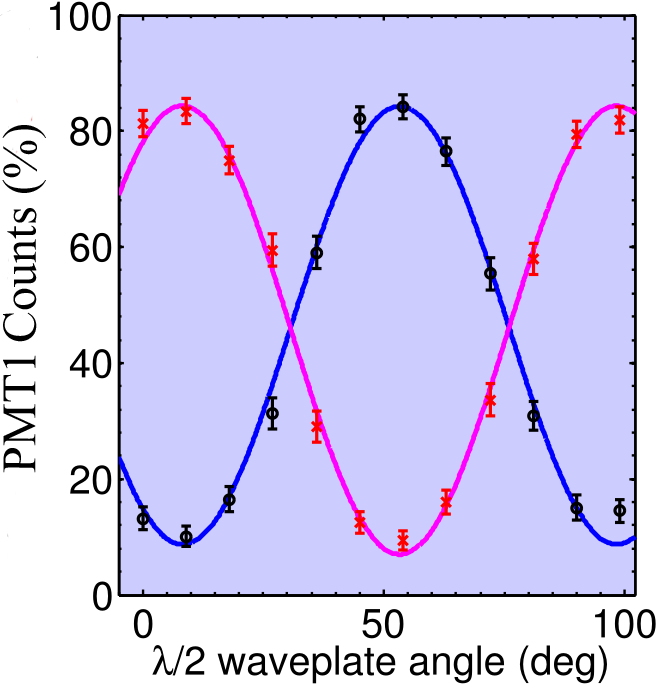} &
(b)\includegraphics[scale = 0.15]{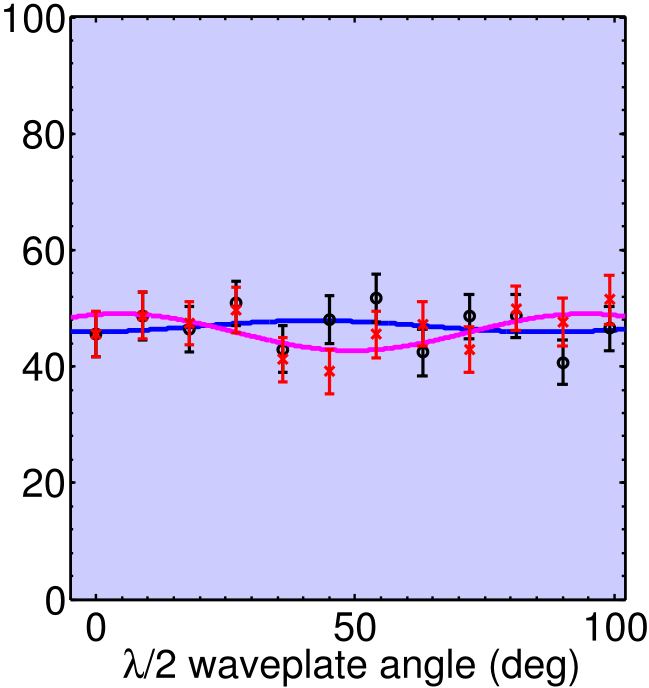} \\
(c)\includegraphics[scale = 0.173]{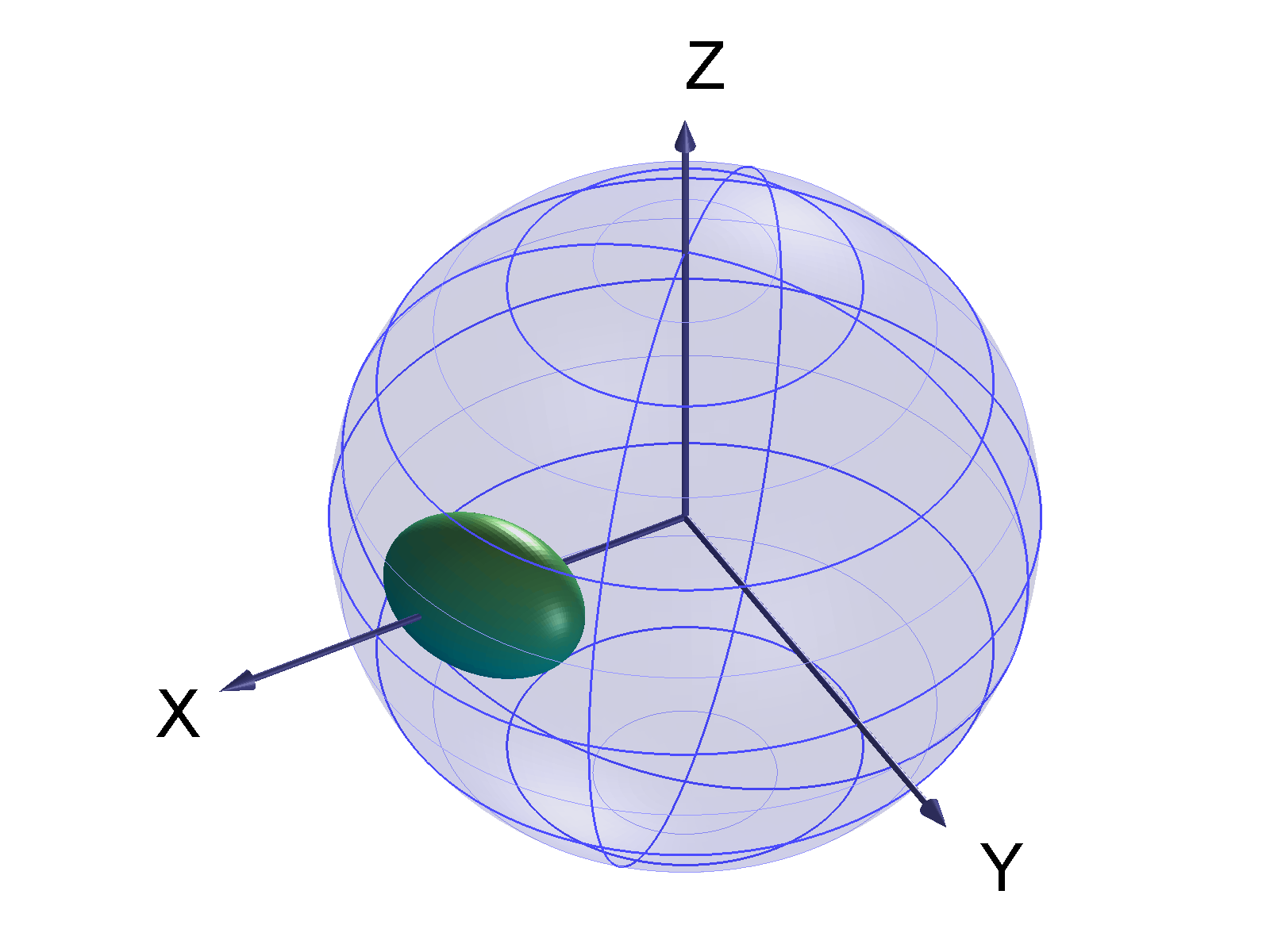} &
(d)\includegraphics[scale = 0.173]{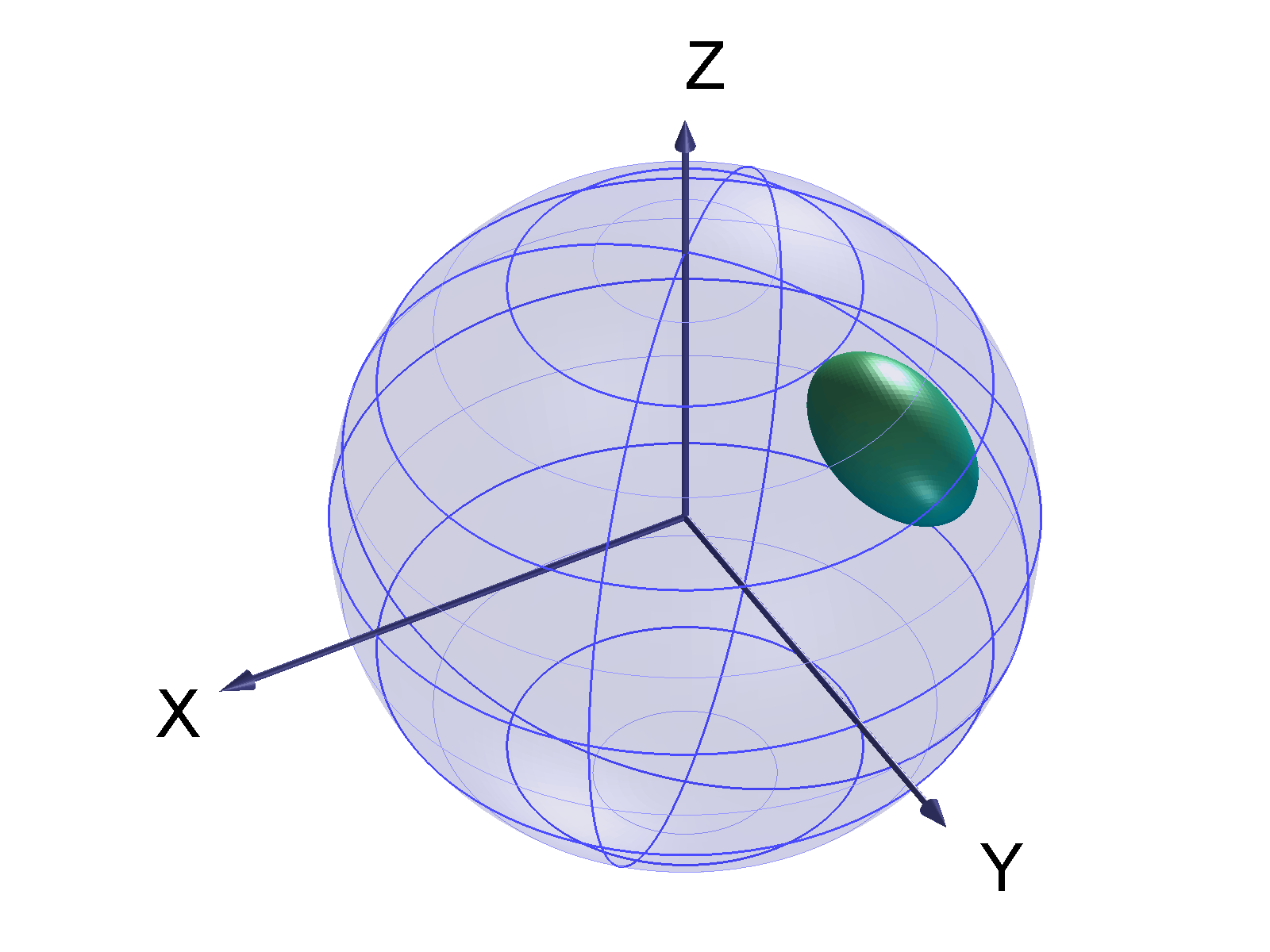}
\end{tabular}
\caption{Spin measurement via photon polarization detection. Quarter-wave retardation plate is oriented such that the circular polarization of scattered photons is transformed into linear. The figures abscissa is the angle of the half-wave retardation plate, that is subsequently rotating the photon polarization direction with respect to that of the PBS. (a) The probability of detecting photons on one port of the PBS vs. the half-wave plate angle, when the spin is initialized to the $|\hat{x}\rangle$ (red filled circles) or $|-\hat{x}\rangle$ (black filled circles) pointer states. Solid lines are sinusoidal fit to the data. As seen, these spin states scatter circularly polarized photons. Furthermore, when the half-wave plate orients the photon polarization detection with the PBS basis, the two spin states can be discriminated with 75\% fidelity. (b) Same as (a), only with the spin initialized in the $|\pm Y \rangle$ states. Starting in this spin state, the spin and photon become entangled. Photons arriving to the PBS therefore carry a statistical mixture of polarization states. The probability of a photon to be measured on a given port of the PBS is hence independent of rotations. Alternatively we performed QPT conditioned on the detection of right or left circularly polarized photons. The surface onto which the Bloch sphere collapses in these two cases is shown in (c) and (d), respectively. As expected, the Bloch sphere collapsed on the $|\hat{x}\rangle$ and $|-\hat{x}\rangle$ states.}\label{PMTClicking}
\end{center}
\end{figure}
 Figure 2(a) plots the absolute value of the entries of the reconstructed process matrix. It is written using the basis elements $|\hat{x} \rangle \langle \hat{x}|=\left(I+\sigma_x\right)/2$, $|-\hat{x} \rangle \langle -\hat{x}|=\left(I-\sigma_x\right)/2$ (projections on the $|\pm \hat{x} \rangle$ states), $-i\sigma_{y}$ and $\sigma_{z}$. In agreement with Eq. \ref{X}, the process matrix is found to be mostly composed of nearly equal contributions of projections on the $|+\hat{x}\rangle$ and $|-\hat{x}\rangle$ states.

 The reconstructed process matrix can be used to evaluate all the post-scattering spin states. Geometrically these states are represented by a surface onto which the Bloch sphere collapses, shown in Fig. 2(b) and Fig. 2(c). An elongated ellipsoid clearly marks the emergence of a spin measurement basis. Here we have chosen the direction along which the pointer basis emerges to be the $\hat{x}$ axis, thus coinciding this coordinate system with the lab coordinate system, shown in Fig. 1. The 1:11 aspect ratio between the spheroid length along the emergent basis and its radial size is dictated by the finite local oscillator phase interval in our data, the finite N.A. of our photon collection lens and quantum projection noise. Further discussion on the resulting spheroid characteristics is given in the supplementary material.

The amount of spin decoherence due to photon scattering can be quantified by calculating the von Neumann entropy of post-scattering states; $S(\rho)=-$Tr$[\rho \ln(\rho)]$. Furthermore, in decoherence theory, a criteria for pointer states is that their entropy does not increase due to their coupling to the environment. We calculated the entropy of all reconstructed final states. Figure 2(d) shows a color map of this final-state entropy plotted over the Bloch sphere of the corresponding initial states. As seen in the figure, spin states along the $\hat{x}$ direction indeed experience the minimal increase in entropy.

% add theoretical transparent bars to the process matrix. Change the order of the figures to
% fit the process matrix with the ellipsoid figures.

\section{Spin measurement in the pointer-state basis}
In order to prove that the $|\pm \hat{x}\rangle$ states on the poles of the emerging ellipsoid are a measurement basis, one needs to show that these states are classically correlated with the scattered photon polarization. For this purpose, we analyzed the polarization of photons that were scattered by different initial spin states. According to Eq. \ref{X}, the $|\pm \hat{x}\rangle$ states scatter photons with a pure circular polarization, whereas other initial spin states become entangled with the photon polarization. The later case leads to a statistical mixture of photon polarization measurement outcomes.

\begin{center}
\begin{figure}
\includegraphics[scale=0.12]{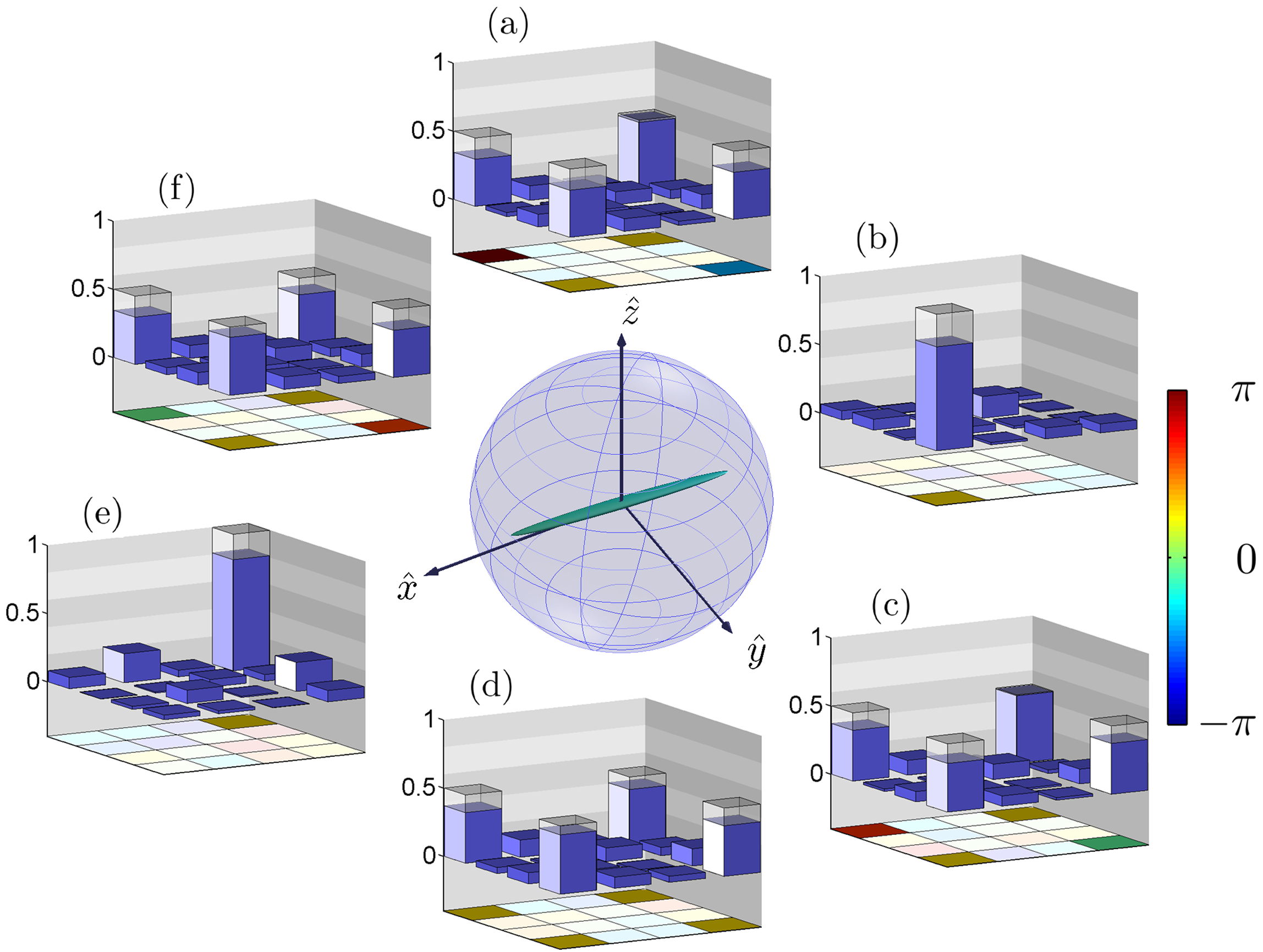}
\caption{State tomography of spin-photon states. Centered around the collapsed spin ellipsoid are six reconstructed spin-photon density matrices. The different density matrices are post-scattering states where the spin was initialized in different initial direction.
The different spin initial states are (starting from the top and clockwise)
$|\uparrow \rangle$, $|-\hat{x}\rangle$, $|\hat{y}\rangle$, $|\downarrow\rangle$,$|\hat{x}\rangle$ and $|-\hat{y}\rangle$. The density matrices are
written in the basis of product states of $| \pm\hat{x}\rangle$ and $|E_{\pm}\rangle$ spin and polarization states. The solid bars are absolute values of entries of the reconstructed density matrices whereas the transparent bars correspond to the values predicted by Eq.\ref{X}. The phases of the different entries are represented by different colors bellow the bars and according to the color map on the right. As seen, the $|\pm\hat{x}\rangle$ pointer states have a single significant matrix element and therefore represent unentangled states. States orthogonal to the pointer-state basis are seen to be highly entangled with the scattered photon polarization.} \label{density}
\end{figure}
\end{center}
Figure 3 shows the results of our photon polarization measurements. Here, we aligned the quarter-wave retardation plate such that it transformed the $| E_{\pm} \rangle$ states to an orthogonal linear basis. The proceeding half-wave plate rotated this linear polarization with respect to the PBS  basis.  Figure 3(a) shows the probability of photon detection on a given port of the PBS vs. the half-wave plate rotation angle. Here, the spin is initialized to the $| \hat{x}\rangle$ (red filled circles) or $| -\hat{x}\rangle$ (black filled circles) states. As expected from a pure polarization state, this probability sinusoidally oscillates as the polarization is rotated. The blue and magenta solid lines are a sinusoidal fit to our data. Furthermore, whenever the two wave-plates transform the emitted circular polarization to match the PBS basis, a clear correlation between the measured polarization and the initial spin state is observed. For example, at a half-wave plate rotation angle of 57 degrees, indicated by the vertical dashed line, a 0.75 correlation between the PMT on which the photon was detected and the initial spin direction is found. Figure 3(b) presents similar data with the spin initialized to the $|\pm \hat{y}\rangle$ states. As expected from a fully mixed polarization state, the photon detection probability on a given PBS port is independent of polarization rotation indicating a lack of classical correlation between the photon polarization state and the initial spin state. An alternative measurement is shown in which QPT was performed conditioned on the detection of right or left circularly polarized photons is shown in Fig. 3(c) and (d), respectively. It can be seen that on detection of right (left) circularly-polarized photon the initial Bloch sphere indeed collapsed to the $|\hat{x}\rangle$ ($|-\hat{x}\rangle$) state.

\begin{center}
\begin{figure}
\includegraphics[scale=0.22]{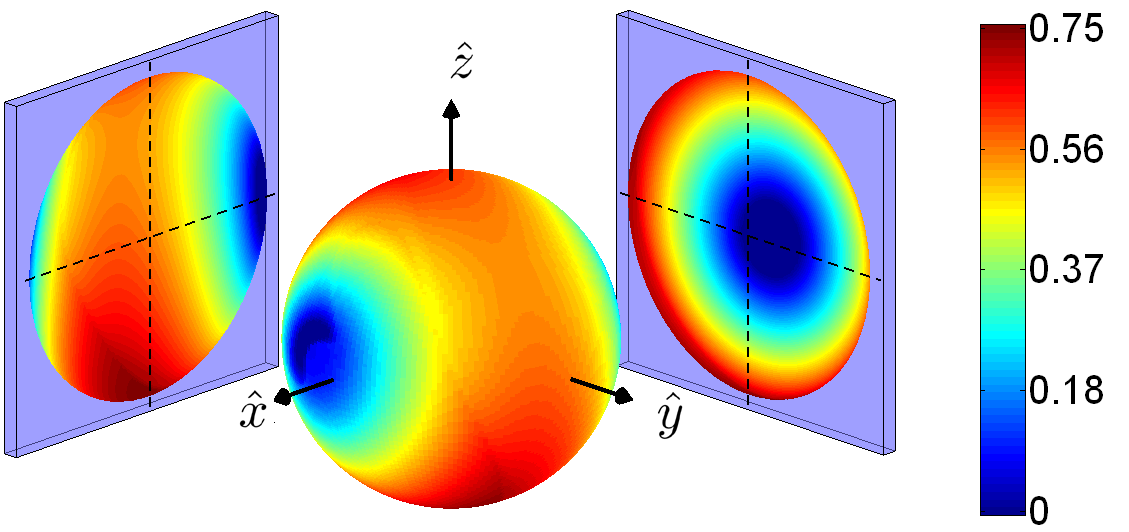}
\caption{A color map of the measured concurrence of every post-scattering spin-photon state. The map is plotted on the Bloch sphere of all initial spin directions. Two projections of the sphere on the $\hat{y}\hat{z}$ and $\hat{z}\hat{x}$ planes are shown as well. Pointer states along the $\hat{x}$ direction are seen to be minimally entangled with the scattered photon polarization.} \label{concurrence}
\end{figure}
\end{center}

\section{Atom-photon entanglement}
\hspace{1pt}
 Starting with an initial spin state other than $| \pm \hat{x}\rangle$ we have seen that both the spin state and the photon polarization-state decohere into statistical mixtures. This decoherence is the result of spin-photon entanglement. To observe this entanglement we preformed quantum state tomography of the combined spin-photon state. Figures 4(a)-(f) present the reconstructed spin-photon density matrices for six different initial spin states, $|\uparrow \rangle$,$|\downarrow \rangle$,$|\pm \hat{x} \rangle$,$|\pm \hat{y} \rangle$. The density matrices are plotted using the $|\pm \hat{x}\rangle$ and $|E_\pm\rangle$ states as a basis. As seen, scattering events in which the spin was initialized along $\hat x$ have a single dominant entry on the diagonal and therefore represent approximately separable states. Alternatively, spin states that were initially oriented along the $\hat y$ or $\hat z$ directions, resulted in highly entangled states.

We quantified the amount of atom-photon entanglement using the concurrence entanglement monotone, $C(\rho)$ \cite{Plenio07}. All atom-photon final density matrices were evaluated by linear combinations of the six reconstructed density matrices. Figure 5 presents a color map of the calculated concurrence values plotted on the Bloch sphere of initial spin states. As seen, the minimum entanglement (C $<$ 0.03) is along the $\hat{x}$ direction. The concurrence monotonically increases up to its maximum value of roughly 0.7 along the sphere circumference in the $\hat{y}\hat{z}$ plane, in consistence with the observed entropy increase shown in Fig. 2(d).

% conclusions and summary
\section{Discussion}
In conclusion, we have studied the evolution of the electronic spin, of a single trapped ion, following resonant scattering of a single photon. To investigate the collapse of the Bloch sphere in the lab frame of reference we performed quantum process tomography stroboscopically with the spin Larmor precession. We found that the Bloch sphere collapses on a basis aligned with the scattered photon propagation direction. In other words, the back-action of observing the atomic spin with light aligns it with the observation direction. We were also able to show that spin states that were aligned with the scattered photon $\hat{k}$ vector, at the moment the photon scattered, emitted photons with a pure circular polarization, whereas any other spin states emitted a statistical mixture of photon polarizations. Furthermore, by detecting the scattered photon polarization, we were able to measure the ion spin in this basis with a fidelity of 0.75. Lastly we have shown, that atom-photon entanglement is larger the further are spin states from the emergent measurement basis. Albeit, classically correlated, the spin pointer-state basis states and the scattered photon polarization are not entangled.

States that are invariant under coupling to the environment are of interest, not only due to their importance in the quantum measurement process, but also due to their potential use for quantum control purposes. Envariant states can span decohernce-free subspaces in which quantum information can be protected \cite{Lidar98}. States that are invariant under dissipation, often refereed to as ``dark states'' act as fixed points in Hilbert space. It was recently suggested that by dark state engineering, in a high dimensional Hilbert space, dissipation can be used to perform quantum computation \cite{Cirac09}. In a  recent experiment, an entangled dark state was engineered using four trapped-ion qubits \cite{Blatt10}. It would be therefore interesting, in the future,  to expand the work presented here and search for states that are invariant under photon scattering by larger arrays of trapped-ion spin qubits.

\begin{acknowledgments}
We thank Dan Stamper-Kurn for useful comments on the manuscript. We gratefully acknowledge the support by the Israeli Science foundations, the Minerva foundation, the German-Israeli foundation for scientific research, the Crown photonics center and Martin Kushner Schnur, Mexico.
\end{acknowledgments}

\appendix
\input{SupplementaryInformationForArXive}

\end{document}

%% file: SupplementaryInformationForArXive.tex
\section{Single Photon Emission or Absorption}
The $^{88}$Sr$^+$ isotope of Strontium has no nuclear magnetic moment and hence there is no hyperfine structure to electronic energy levels.  A qubit is encoded in the two electronic-spin states in the $5s^{2}S_{1/2}$ ground level. Using a weak, linearly polarized, laser beam at a wavelength of 422 nm, photons were scattered on the $5s^{2}S_{1/2} \rightarrow 5p^{2}P_{1/2}$ transition. The process of laser excitation of the electron, followed by spontaneous photon emission, returns the electron to the two-fold ground-state manifold. In terms of spin evolution, spontaneous photon scattering can therefore formally be described as a non-unitary, yet trace-preserving, operation. When limited to this transition, absorption or emission of a photon maps between the  $S_{1/2}$ and $P_{1/2}$ spin 1/2 manifolds. It is therefore convenient to think of the coupling between these manifolds in terms of spin 1/2 (Pauli) operators.

The electromagnetic field couples the $S_{1/2}$ and $P_{1/2}$ levels via the dipole interaction Hamiltonian, $\textbf{H}_{int} =-e\vec{\textbf{r}}\cdot\vec{\textbf{E}}$, were, $e\vec{\textbf{r}}$
is the electric dipole operator. Here we use the quantized electric field operator:
\begin{equation}
\vec{\textbf{E}}=i\sqrt{\frac{\hbar\omega}{2V\varepsilon_0}}\sum
_{\vec{k},E}\left(\vec{E} e^{i\vec{k}\vec{R}}\otimes
\textbf{a}_{\vec{k},\vec{E}} -\vec{E}^{*}
e^{-i\vec{k}\vec{R}}\otimes
\textbf{a}^{\dagger}_{\vec{k},\vec{E}}\right),
\end{equation}
where $\textbf{a}_{\vec{k},\vec{E}}$ and
$\textbf{a}^{\dagger}_{\vec{k},\vec{E}}$ are photon
annihilation and creation operators, respectively, $\vec{E}$ is the
photon polarization and $\vec{k}$ is its wave-vector. Here $\vec{R}$ is the atomic center-of-mass location which determines the electromagnetic field phase, and is considered here as a classical coordinate rather than a quantum operator. Electron decay is followed by photon emission,
\begin{equation}
\textbf{H}_{\vec{k},emm}=ie\sqrt{\frac{\hbar\omega}{2V\varepsilon_0}}\sum
_{E}\left(\vec{\textbf{r}}\cdot\vec{E}^{*}\right)
e^{-i\vec{k}\vec{R}}\otimes \textbf{a}^{\dagger}_{\vec{k},\vec{E}}.
\end{equation}
Note that $\textbf{a}^{\dagger}_{\vec{k},\vec{E}}$
acts in the photonic part of Hilbert space while $\vec{\textbf{r}}$ acts on the electronic part of Hilbert space, which is a tensor product of the electron motion and spin. The product
$\vec{\textbf{r}}\cdot\vec{E}^*$ can be written using spherical
coordinates $\vec{\textbf{r}}\cdot\vec{E}^* =\sum_q \textbf{r}_q^*
E^*_q$, where $q=\{-1,0,1\}$. Using the Wigner-Eckart theorem the operator components $\textbf{r}_q$ can be reduced to
\begin{equation}
\langle j',m'|\textbf{r}_q|j,m\rangle = \frac{\langle l' || \textbf{r}||l\rangle}{\sqrt{2j'+1}}\langle l,1;q,m|l',m'\rangle,
\end{equation}
 where, $l$, $s$ and $j$ are the orbital, spin and total angular momentum numbers, respectively. The primed variables stand for the excited states while the unprimed variables stand for the ground states. The term $\langle l,1;q,m|l',m'\rangle$ is the Clebsch - Gordan coefficient. In this work $l=l'=0$ and $j=j'=1/2$, and it can be shown that the operation of the different $\textbf{r}$ components in the spin part of Hilbert space can be written using the Pauli matrices:
\begin{align*}
\textbf{r}_0 &= \sqrt{\frac{1}{2}}\sqrt{\frac{1}{3}}\langle S_{1/2} || \textbf{r}||P_{1/2}\rangle\sigma_z, \\
\quad \textbf{r}_{\mp1}&=
\mp\sqrt{\frac{1}{2}}\sqrt{\frac{2}{3}}\langle S_{1/2} || \textbf{r}||P_{1/2}\rangle\sigma^\pm= \\
&\mp\frac{1}{2}\sqrt{\frac{1}{2}}\sqrt{\frac{2}{3}}\langle S_{1/2} || \textbf{r}||P_{1/2}\rangle\left(\sigma_x\pm
i\sigma_y\right).
\end{align*}

The electric field polarization components can, correspondingly, be written in
cartesian coordinates:
\begin{displaymath}
\hat{E}_0 = \hat{E}_z, \qquad \hat{E}_{\pm1} =
\frac{1}{\sqrt{2}}\left(\hat{E}_x\pm i \hat{E}_y\right).
\end{displaymath}
Plugging this to the emission Hamiltonian, at
$\vec{R}=0$ and for some fixed $\vec{k}$ gives:
\begin{widetext}
\begin{equation}
\textbf{H}_{\vec{k},emm}=ie\sqrt{\frac{\hbar\omega}{2V\varepsilon_0}}\langle S_{1/2} || \textbf{r}||P_{1/2}\rangle\left[
\left(\vec{\sigma}\cdot\vec{E}^{*}_1\right)\otimes
\textbf{a}^{\dagger}_{\vec{k},\vec{E}_1}+\left(\vec{\sigma}\cdot\vec{E}^{*}_2\right)\otimes
\textbf{a}^{\dagger}_{\vec{k},\vec{E}_2}\right]
\end{equation}
\end{widetext}
This Hamiltonian implies that emission of a linearly polarized photon by the ion is followed
by a spin rotation around the emitted photon polarization vector. To see this, recall that the results in our experiment are post-selected on the detection of a single emitted photon. Formally, if $\textbf{U}(t)$ is the unitary evolution then at short times $\textbf{U}(t)=\exp(-i\textbf{H}t/ \hbar)\approx 1- i\textbf{H}t/ \hbar\ldots$. By using a weak excitation pulse and conditioning our measurement on a single photon detection event we post-select events generated by a single application of $-i\textbf{H}t/ \hbar$. Photon absorption applies $\vec{\sigma}\cdot \vec{E}_L=e^{i\frac{\pi}{2}(\vec{\sigma}\cdot \vec{E})}$, whereas photon emission applies $(\vec{\sigma}\cdot\vec{E_p} )^\dagger$ where $E_L$ and $E_p$ are the laser and the measured photon polarizations respectively.

\section{Process tomography in the lab frame of reference}
The ion electronic spin preforms Larmor precession around the static magnetic field
direction. In a frame rotating with the spin, the lab axes $\hat{x}$ and
$\hat{y}$ rotate around the $\hat{z}$ axis at the Larmor precession frequency of 3.5 MHz.
The detected photon scattering direction however is fixed by the setup imaging system, in the lab frame. Therefore, to observe the Bloch sphere collapse on a basis aligned with the detected photon $\vec{k}$-vector direction we preformed process tomography, in the lab frame of reference, stroboscopically with the spin precession rather than in a frame rotating with it. A local oscillator, which was synchronized with the spin rotation frequency, triggered the RF pulses with which we preformed spin rotations and was used to generate the pulse. The spin direction within the Larmor precession cycle at the time the photon was scattered was estimated by the local oscillator phase at the
photon time of detection. It is extracted by a time-to-amplitude
converter (TAC) which is periodically reset by the local oscillator zero
crossing, and triggered by the PMT whenever a photon is detected. Figure
\ref{PhotonPhasedistribution} shows the
distribution of local oscillator phases recorded at different photon detection events, spanned between $0$ to $2\pi$ radians. Here $2\pi$ corresponds to a delay of 285 nsec between the local oscillator zero crossing and the photon detection event. Restricting process tomography to events that occurred during a given phase window results in a collapse of the Bloch sphere on an elongated ellipsoid. The lower part of Fig. \ref{PhotonPhasedistribution} shows the different ellipsoids that result from post-selection of different phase windows marked by the different colors. The main axis of the resulting ellipsoids lies in the sphere equatorial ($\hat{x}$-$\hat{y}$) plane. The angle of the ellipsoid main axis from the $\hat{x}$ axis is given by the mean phase in the phase acceptance window.
\begin{widetext}
\begin{center}
\begin{figure}[!htb]
\includegraphics[scale=0.35]{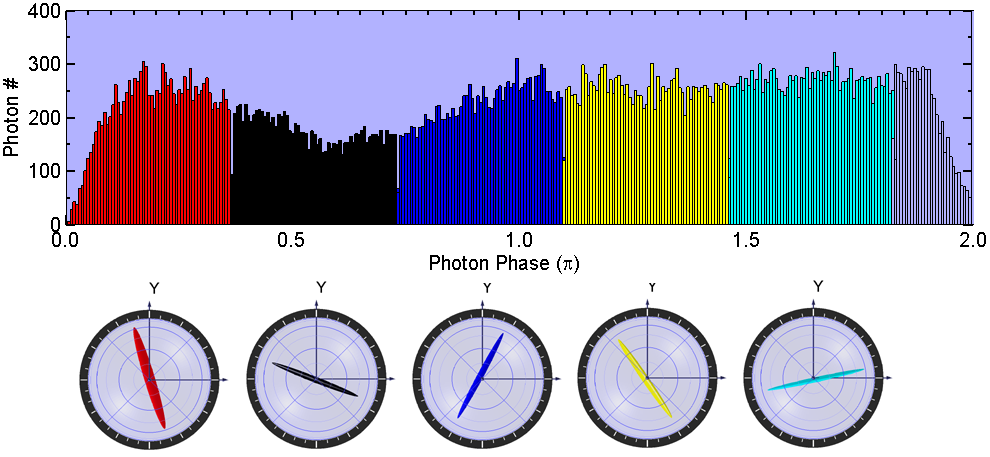}
\caption{The distribution of spin Larmor precession phases at photon detection times. This distribution was
divided to five phase windows marked by the different colors. Preforming QPT using events that were post-selected from each group
yields the corresponding ellipsoids shown on the bottom part of the figure. All ellipsoids lie largely in the x-y plane, and the angle of their main axis from the x direction equals the average phase in the selected window.} \label{PhotonPhasedistribution}
\end{figure}
\end{center}
\end{widetext}
\newpage
The ellipsoid aspect ratio depends on the width of the phase acceptance window. Figure 2 shows the surface on which the Bloch sphere is projected as a function of the phase acceptance window width, which is marked in magenta on the corresponding distributions. As seen, the aspect ratio of the ellipsoid tends to one when the phase acceptance window is large enough. When all phases are accepted the resulting ellipsoid is a disk in the equatorial plane \cite{Akerman2011}. The width of all spheroids in the $\hat{z}$ direction is roughly consistent with projection noise. To model the process matrix following scattering of a single photon we average the ideal process matrix $\chi_{\vec{k}(\theta,\phi)}$ over different photon scattering directions in the rotating frame. This matrix describes the process which is applied on the spin given that a photon was scattered in the $\vec{k}\left(\theta,\phi\right)$ direction. In the $\{I,\sigma_x,-i\sigma_y,\sigma_z \}$ basis \cite{IkeAndMike} it is defined as:
\begin{displaymath}
\chi_{\vec{k} \left(\theta,\phi\right)} = \frac{1}{4}\left[ \begin{array}{c c c c} a & b & c &  0 \\
b^* & d &  e & 0\\
c^* & e^* & f & 0 \\
0 & 0 & 0 & 0  \end{array}\right],
\end{displaymath}
where,
\begin{displaymath}
\begin{array}{l}
a = 2\sin^2(\theta)\\
b = i\sin(2\theta)\sin(\phi)\\
c = -\sin(2\theta)\cos(\phi)\\
d = 2\cos^2(\phi) + 2\cos^2(\theta)\sin^2(\phi)\\
e = i(\sin(2\phi)\cos^2(\theta)-\sin(2\phi))\\
f = 2\cos^2(\phi)\cos^2(\theta) + 2\sin^2(\phi).
\end{array}
\end{displaymath}
Here $\theta$ and $\phi$ are the the scattered-photon $\vec{k}$-vector nutation and precession angles, respectively, relative to the $\hat{z}$ axis, as can be seen in Fig. \ref{LFOR}.
\begin{figure}
\begin{center}
\includegraphics[scale=0.2]{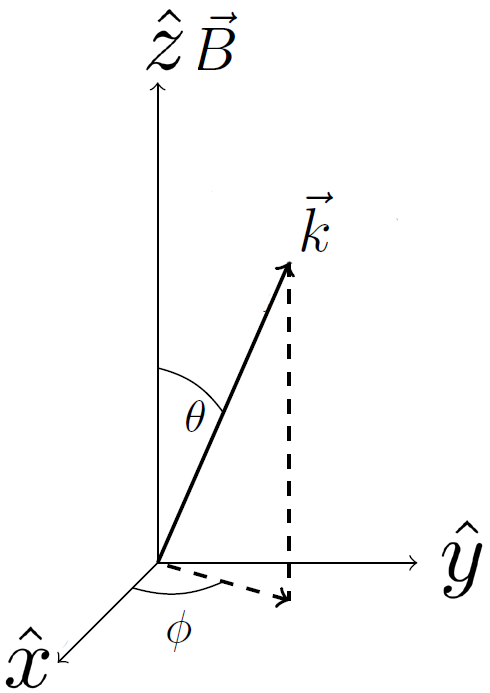}
\caption{$\vec{k}$-vector orientation in the lab frame of reference.} \label{LFOR}
\end{center}
\end{figure}
Photons are collected through our imaging system numerical aperture (NA) of 0.31. This result in a $\vec{k}$-vectors distribution, furthermore, the size of the phase acceptance window [$-\phi_{LO}$:$\phi_{LO}$] determines the width of a square distribution of scattering directions in the rotating frame.
\begin{equation}
\label{chiEq}
\chi_{scatt} = \int_{-\phi_{LO}}^{\phi_{LO}}d\phi'\int_{-\theta_{max}}^{\theta_{max}} d\theta'
\chi_{\vec{k} \left(\theta',\phi'\right)}.
\end{equation}
Here $\theta'$ and $\phi'$ are the spherical angles (starting at
$\hat{x}$ for $\theta'=\pi/2$ and $\phi'=0$) and
$\theta_{max}=\pi/2 +\sin^{-1}\left(NA\right)$ is the maximally possible
scattered photon $\vec{k}$-vector orientation to be detected. Figure \ref{RotatingZurek3} shows the aspect ratio of the ellipsoid in the equatorial plane. Solid line is calculated
using Eq. \ref{chiEq} and red crosses are given by a fit of and ellipsoid surface to the data shown in Fig. \ref{ZeurekEllipsoidChange} (a)-(k). The two are seen to be in a good agreement.
\begin{widetext}
\begin{center}
\begin{figure}[!htb]
\begin{tabular}{|cc|cc|}
 \hline
(a)\includegraphics[scale=0.15]{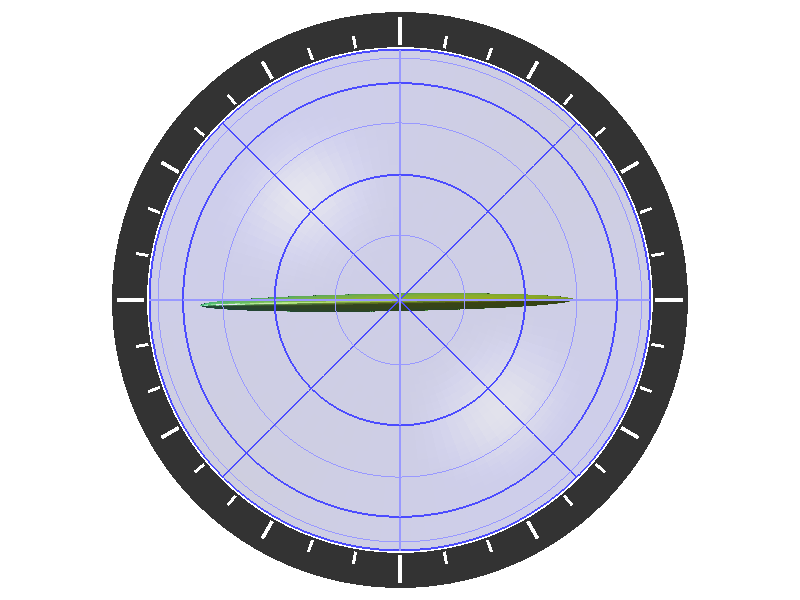} &
\includegraphics[scale=0.15]{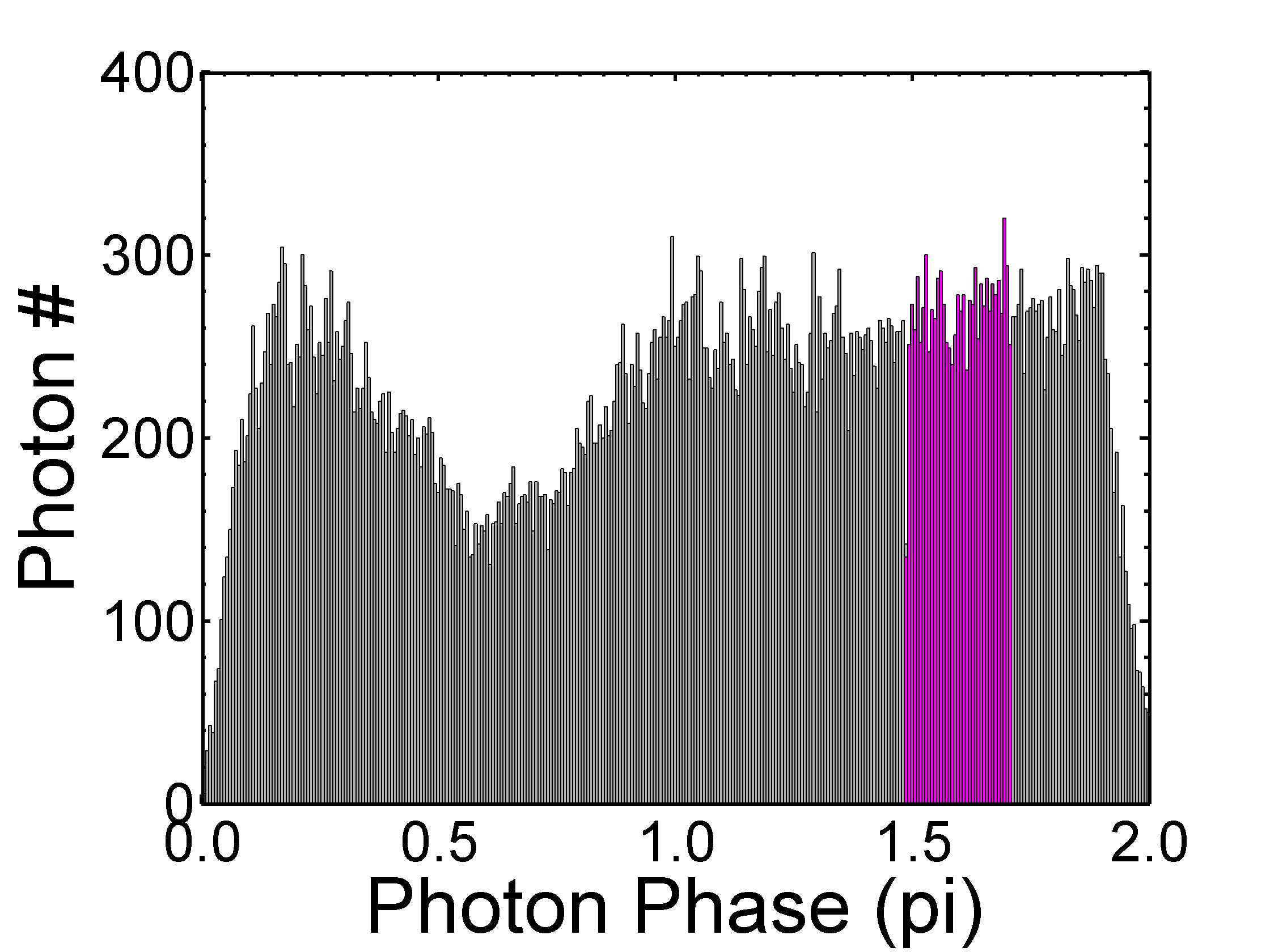} &
(b)\includegraphics[scale=0.15]{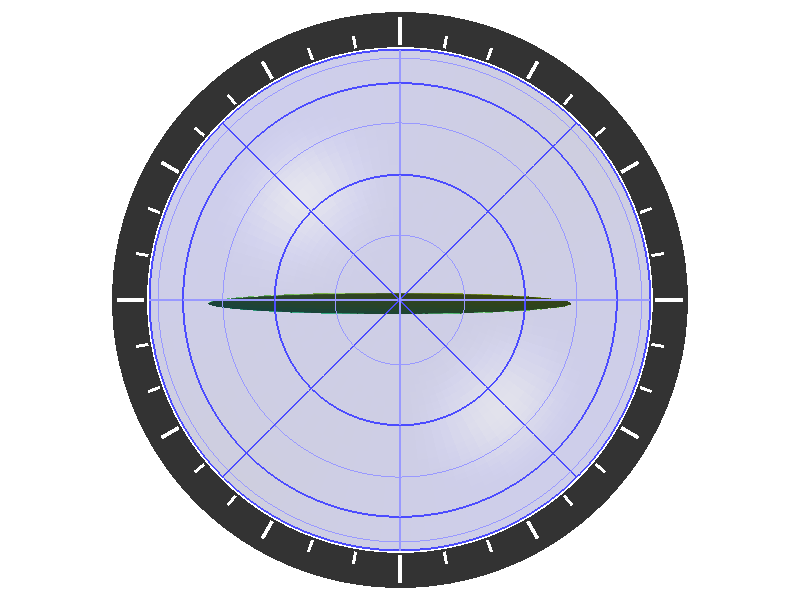} &
\includegraphics[scale=0.15]{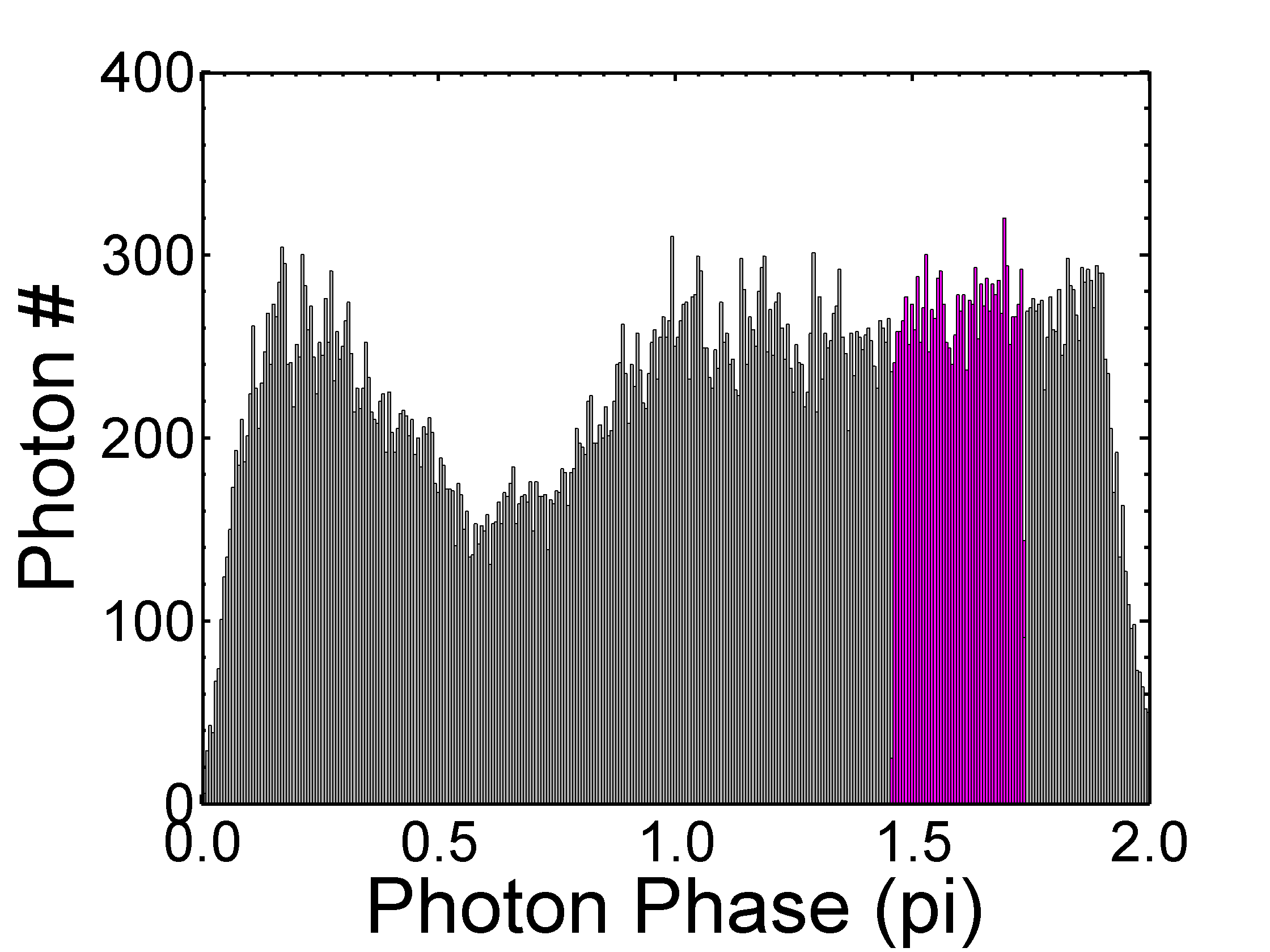} \\ \hline
(c)\includegraphics[scale=0.15]{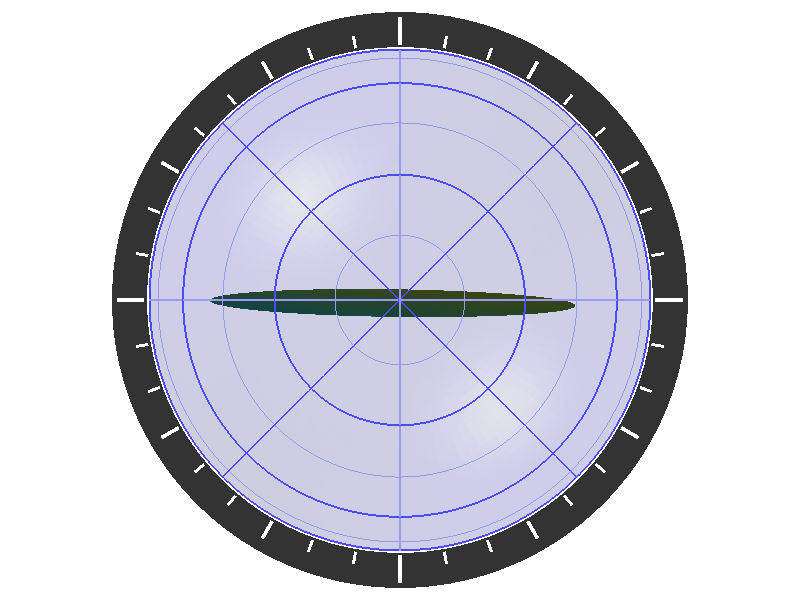} &
\includegraphics[scale=0.15]{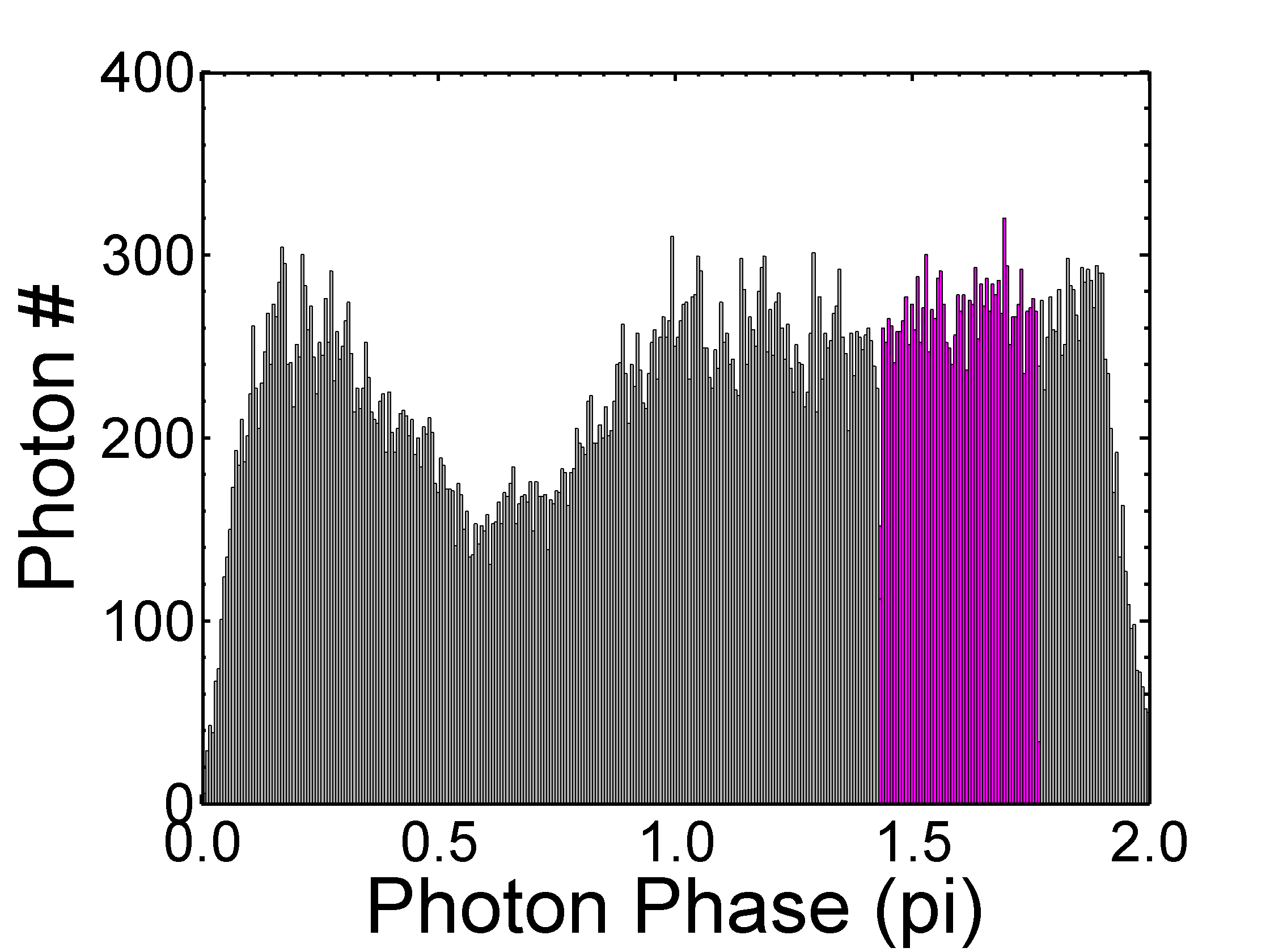} &
(d)\includegraphics[scale=0.15]{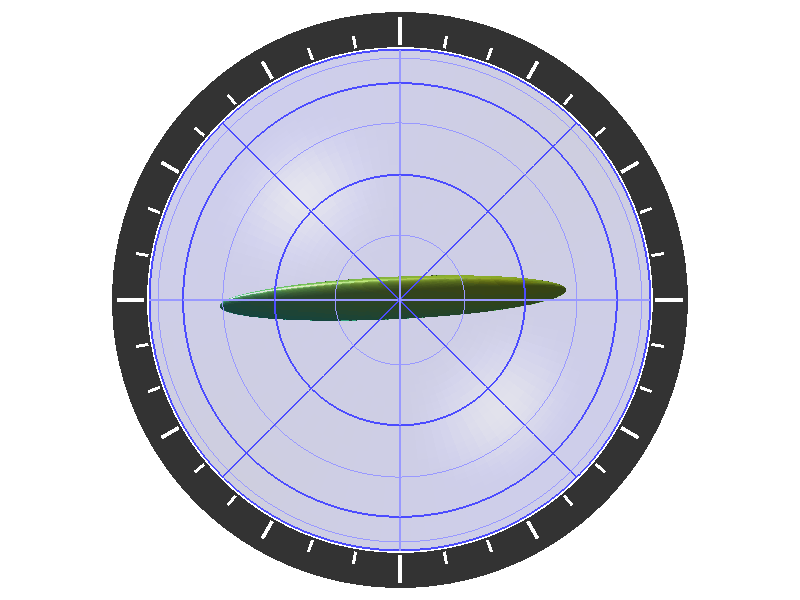} &
\includegraphics[scale=0.15]{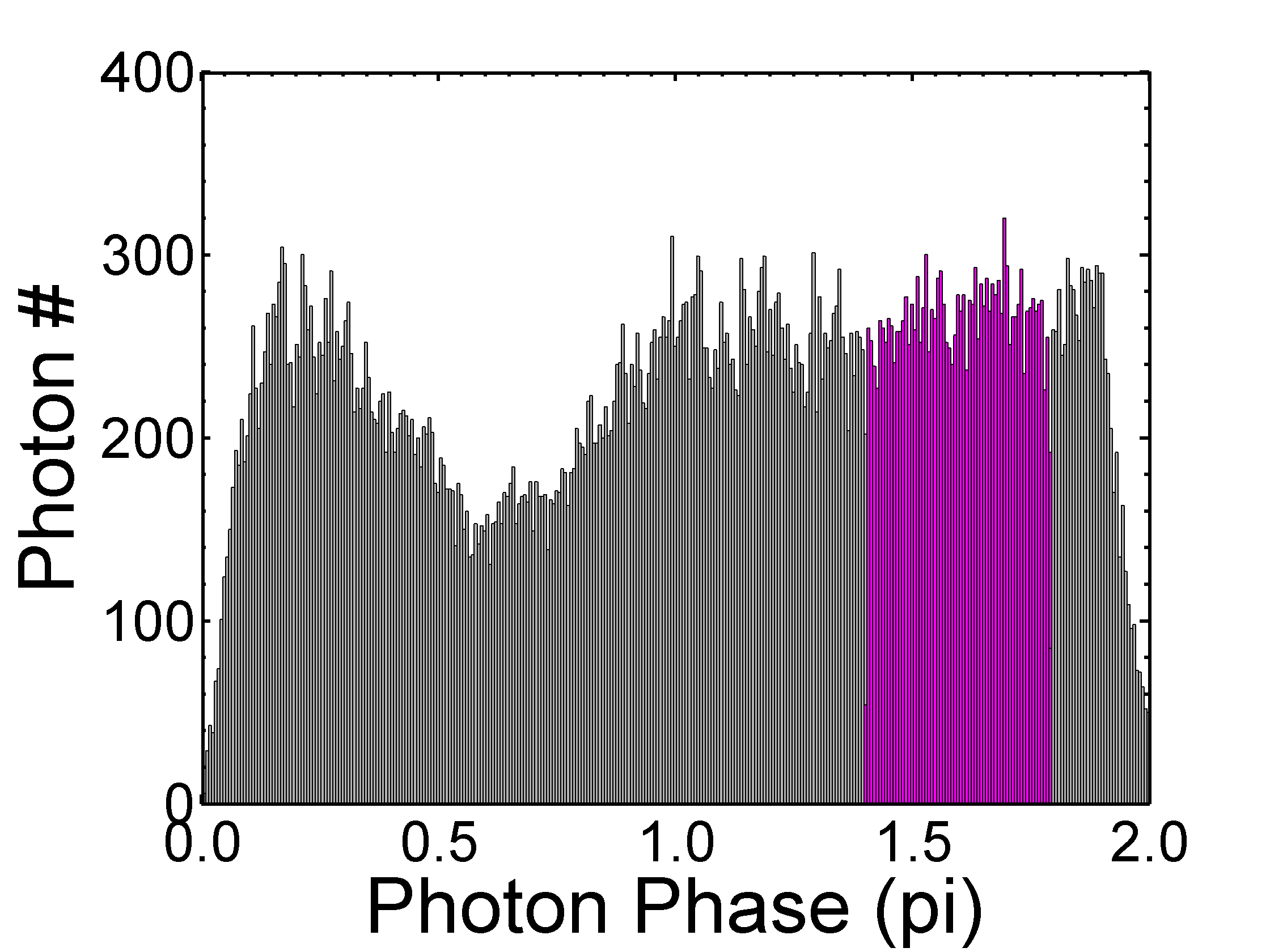} \\ \hline
(e)\includegraphics[scale=0.15]{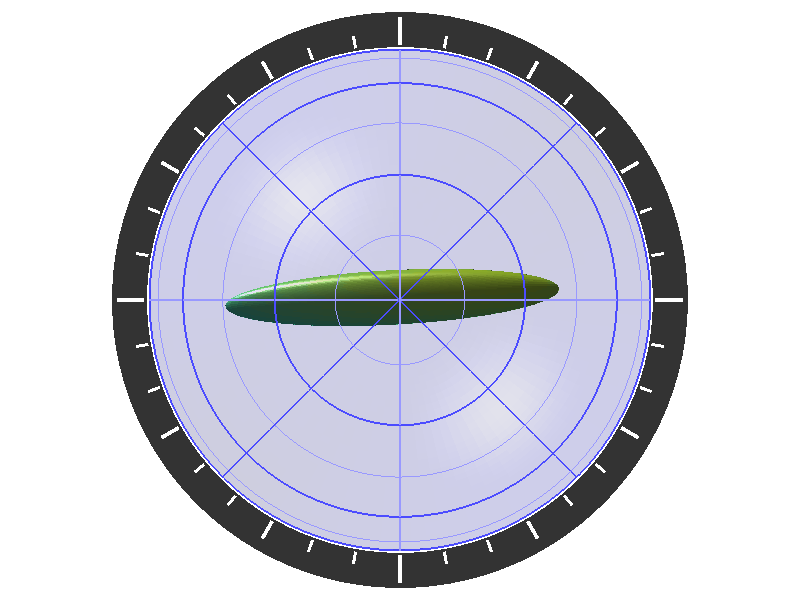} &
\includegraphics[scale=0.15]{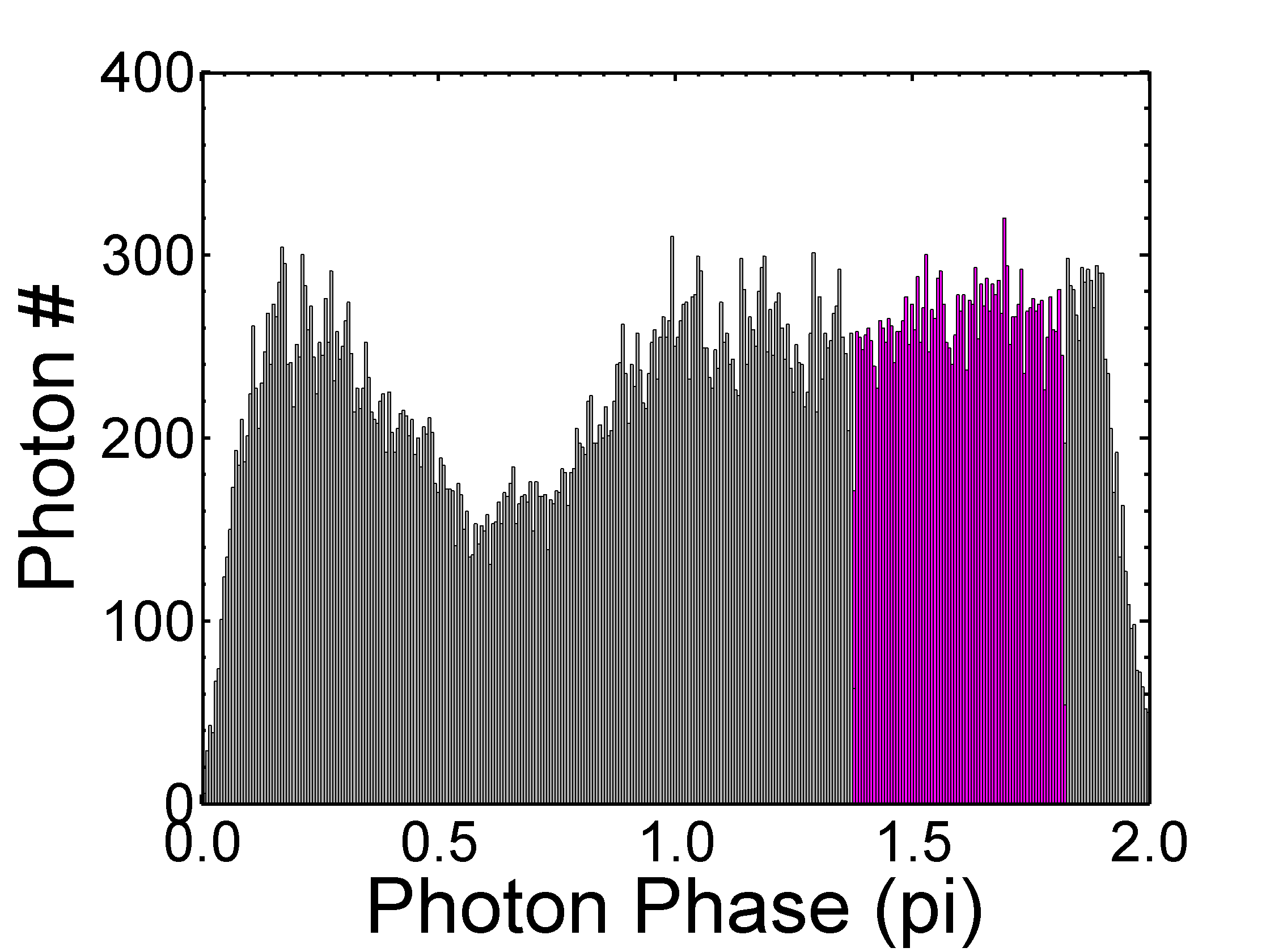} &
(f)\includegraphics[scale=0.15]{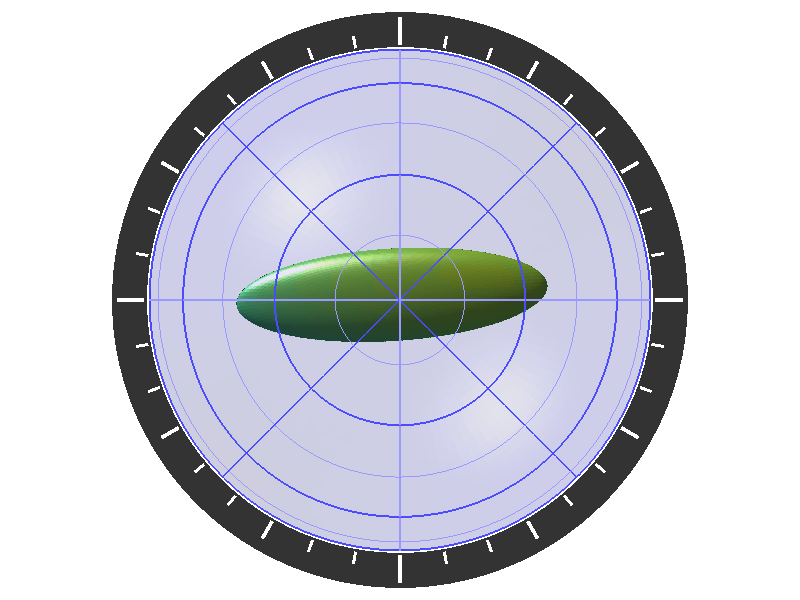} &
\includegraphics[scale=0.15]{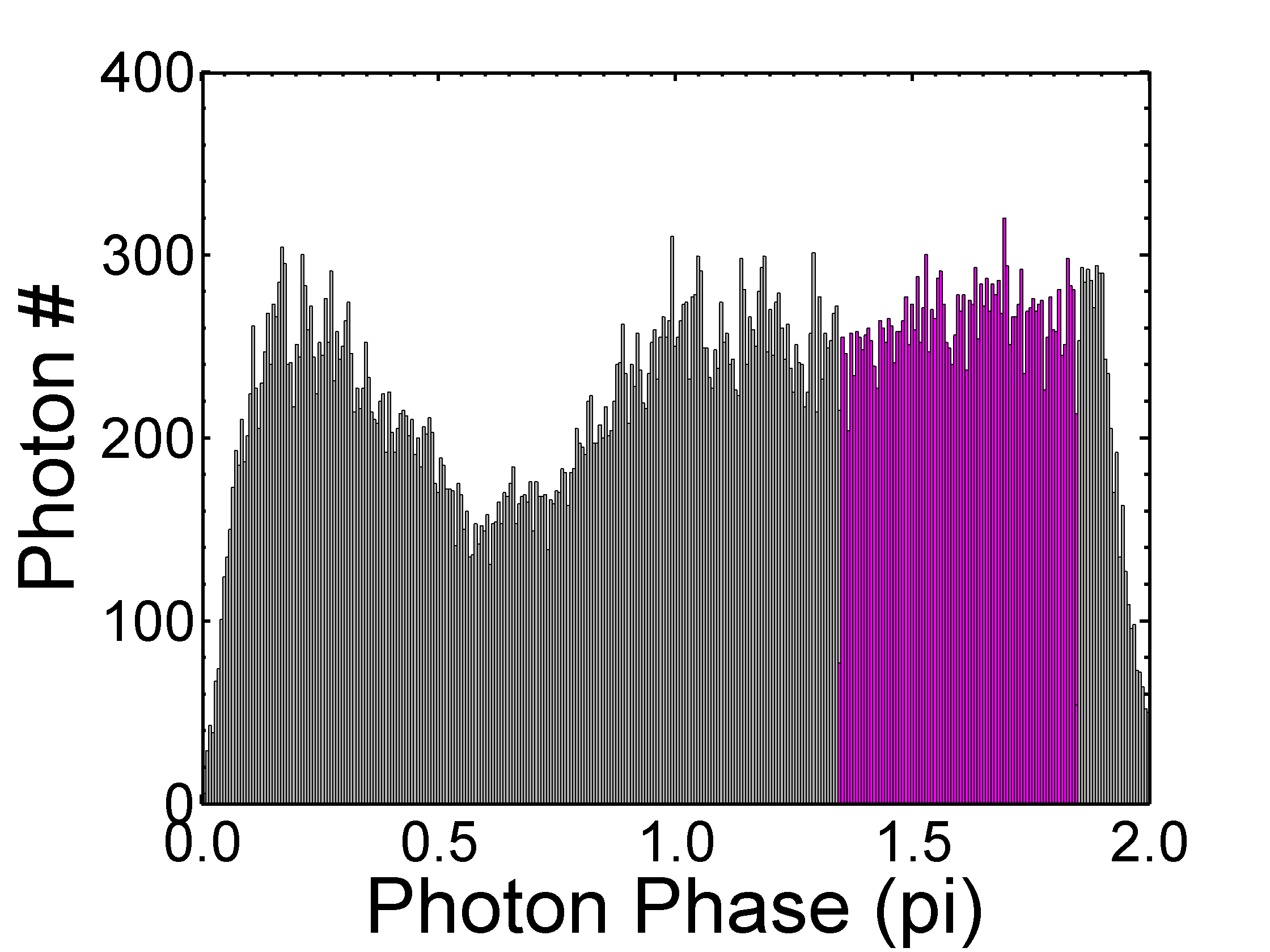} \\ \hline
(g)\includegraphics[scale=0.15]{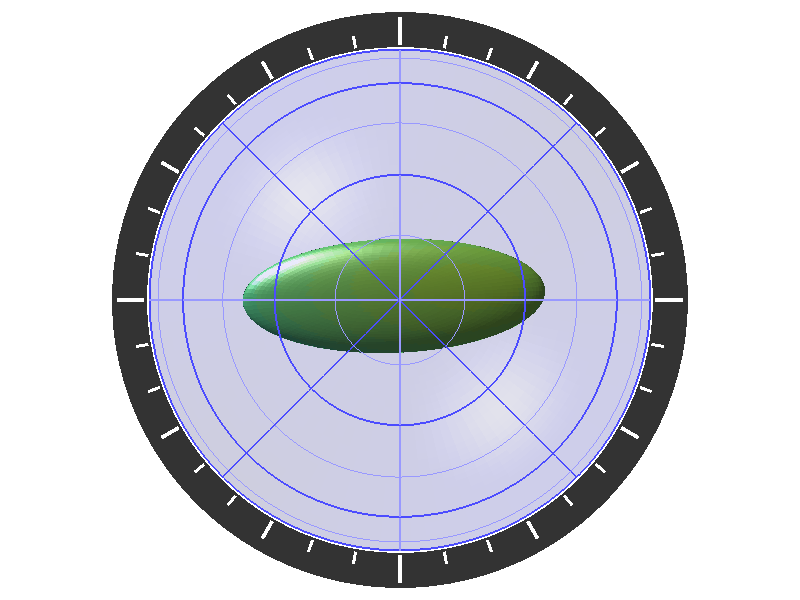} &
\includegraphics[scale=0.15]{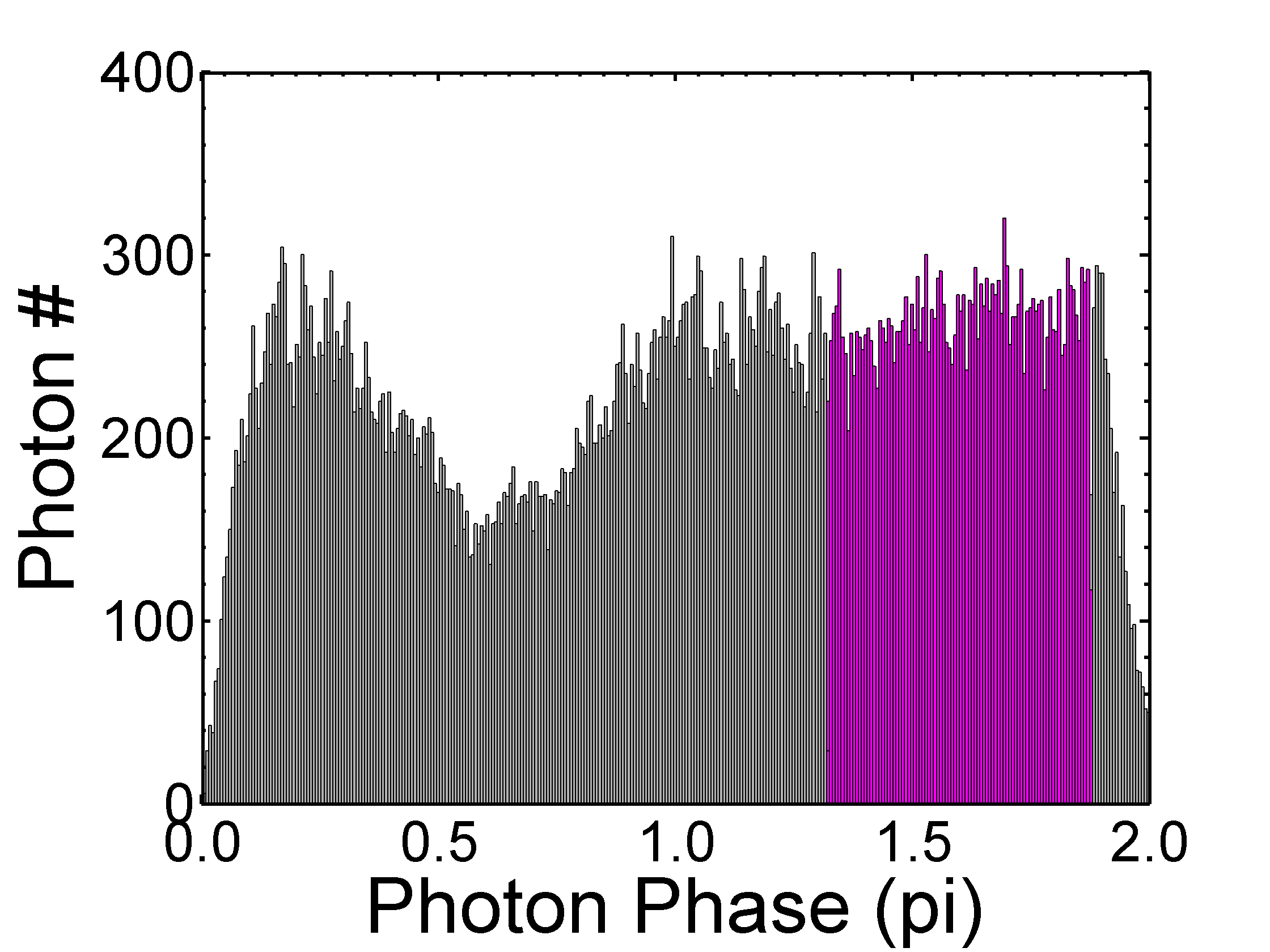} &
(h)\includegraphics[scale=0.15]{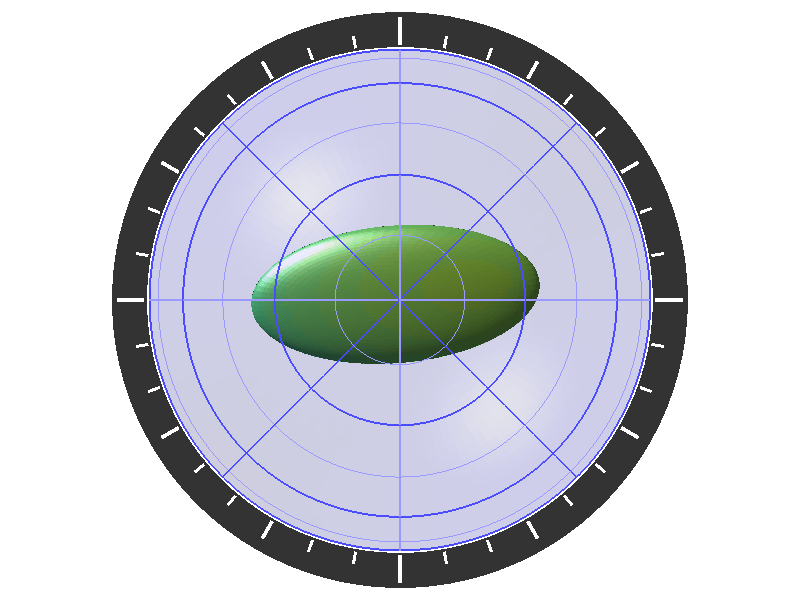} &
\includegraphics[scale=0.15]{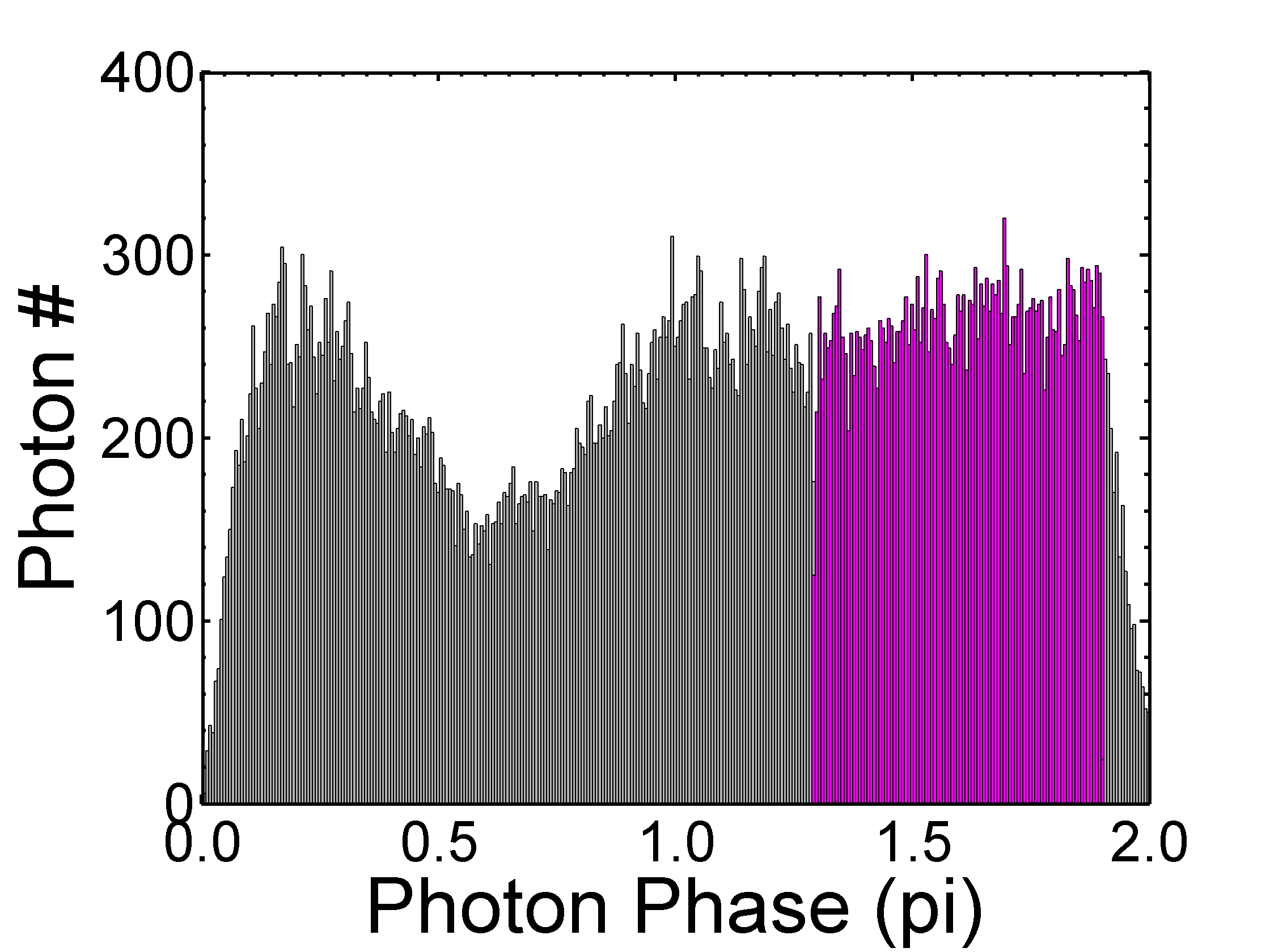} \\ \hline
(i)\includegraphics[scale=0.15]{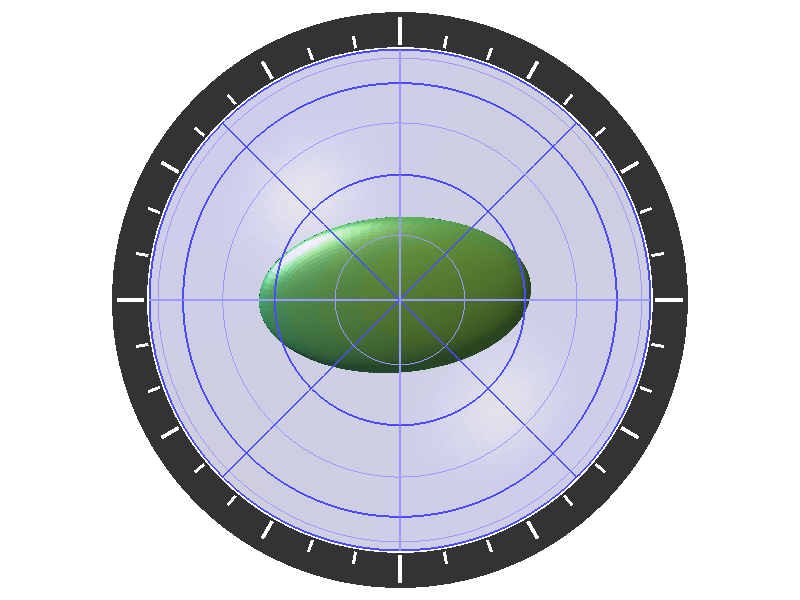} &
\includegraphics[scale=0.15]{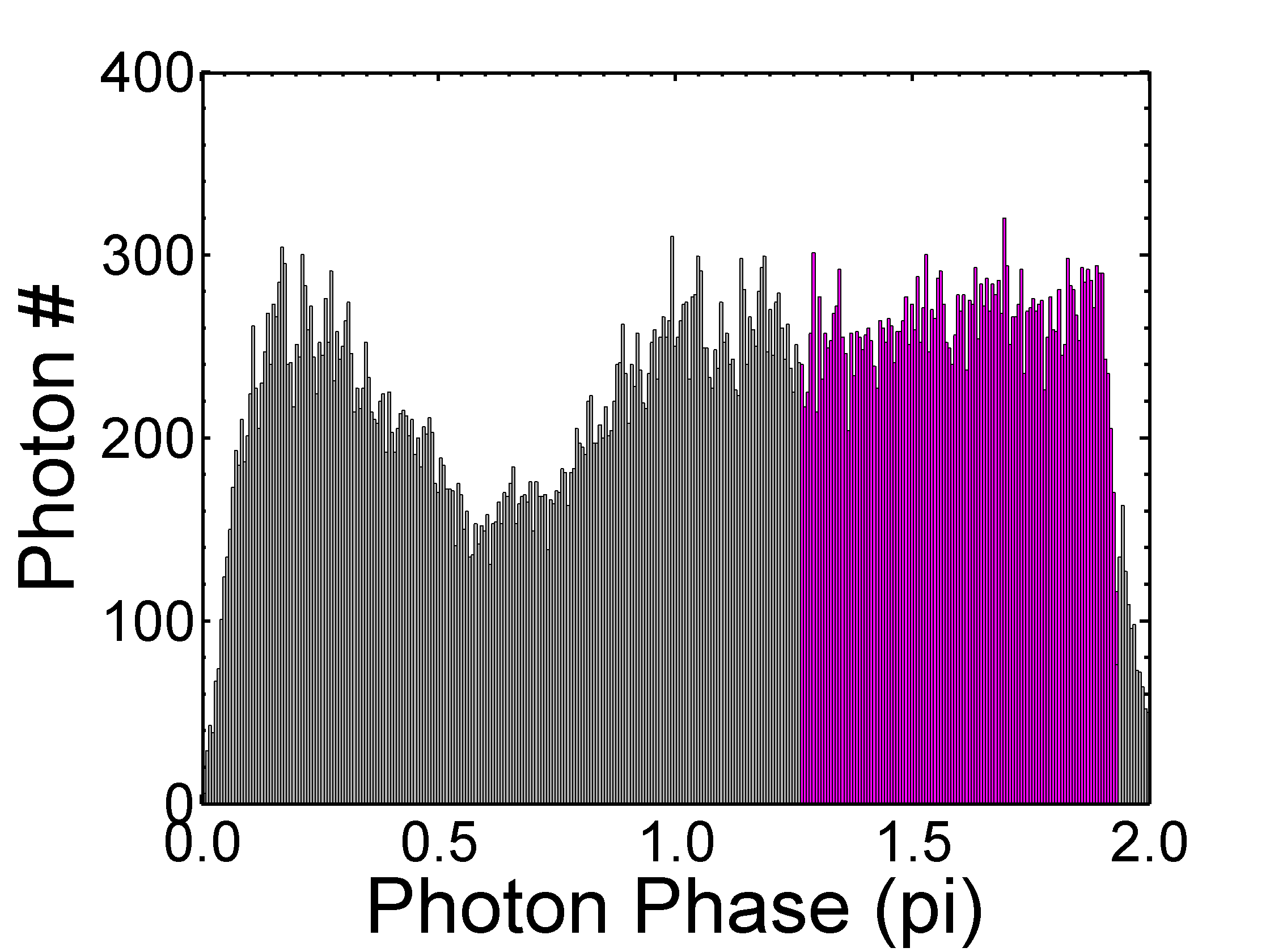} &
(j)\includegraphics[scale=0.15]{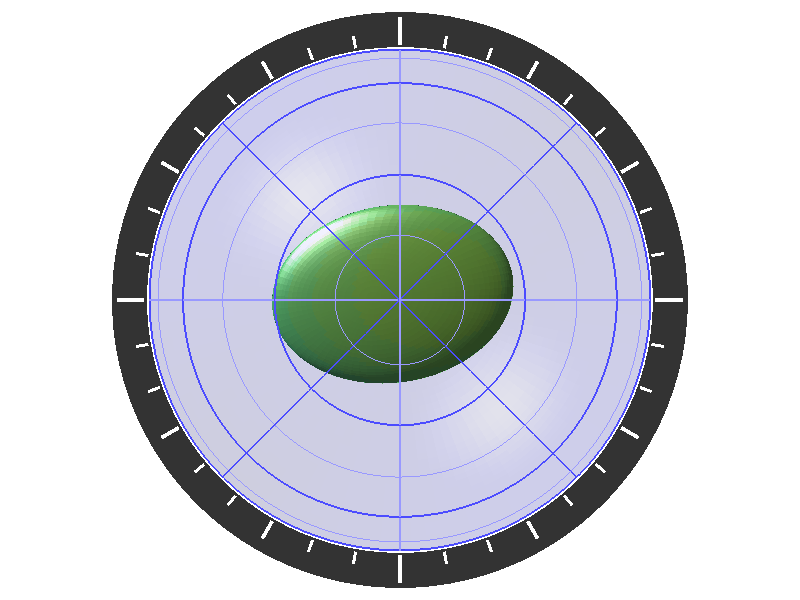} &
\includegraphics[scale=0.15]{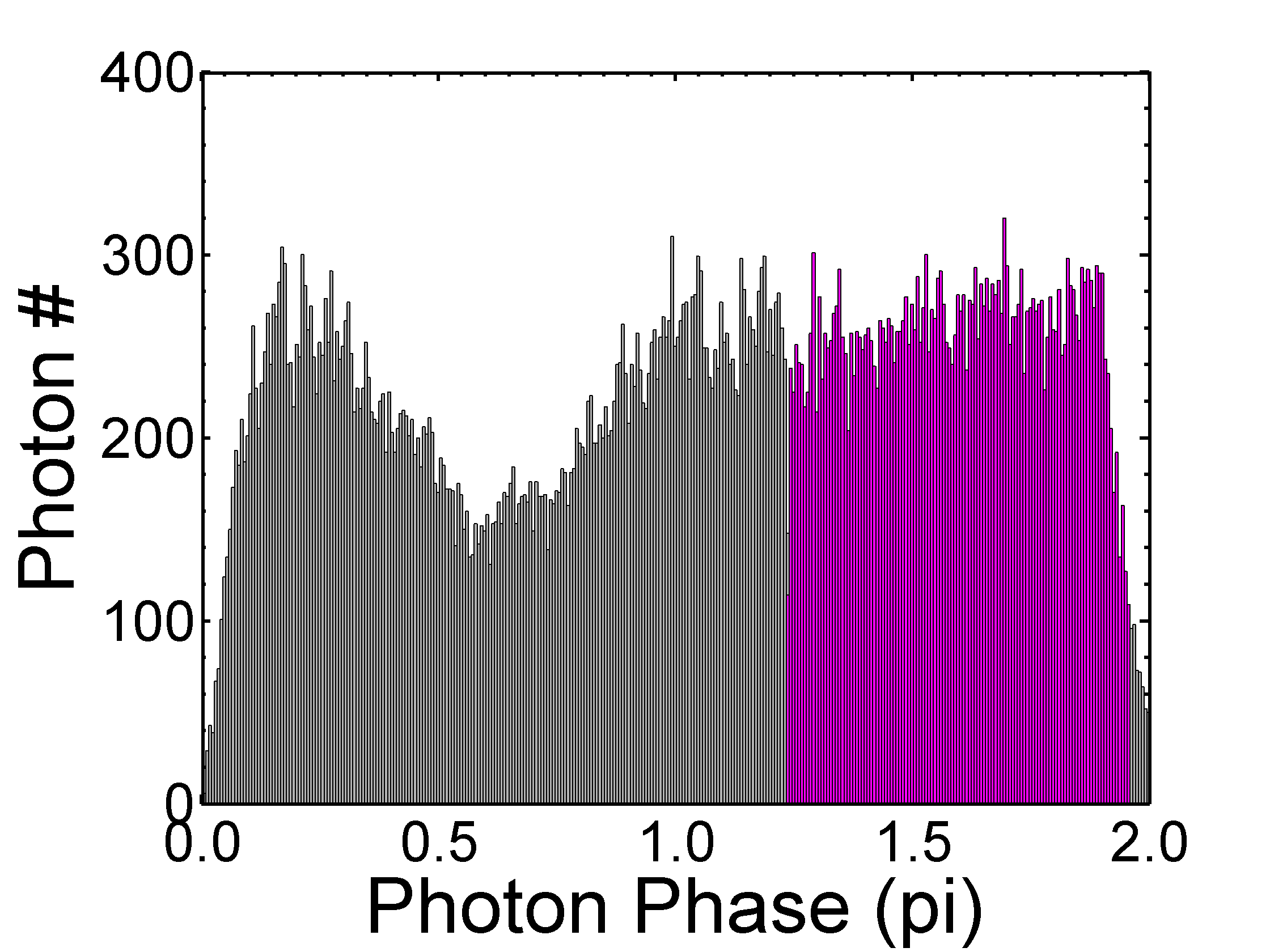} \\ \hline
(k)\includegraphics[scale=0.15]{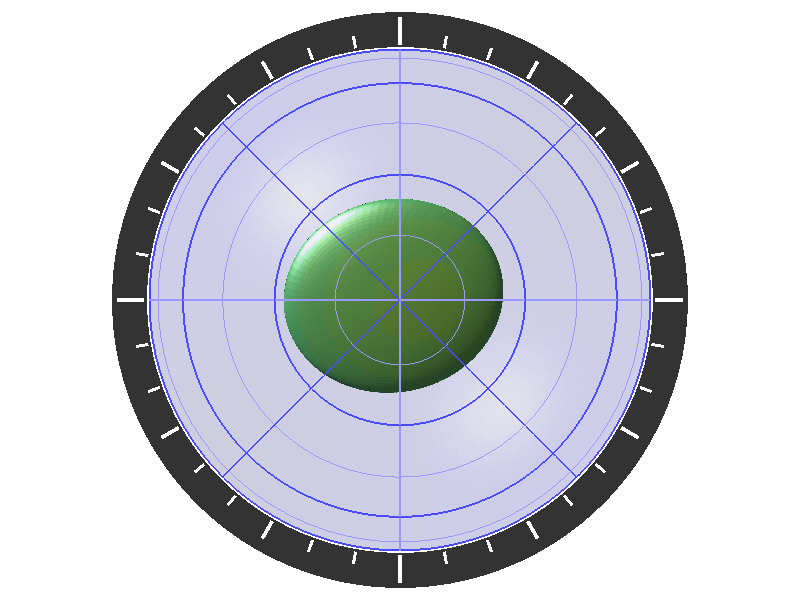} &
\includegraphics[scale=0.15]{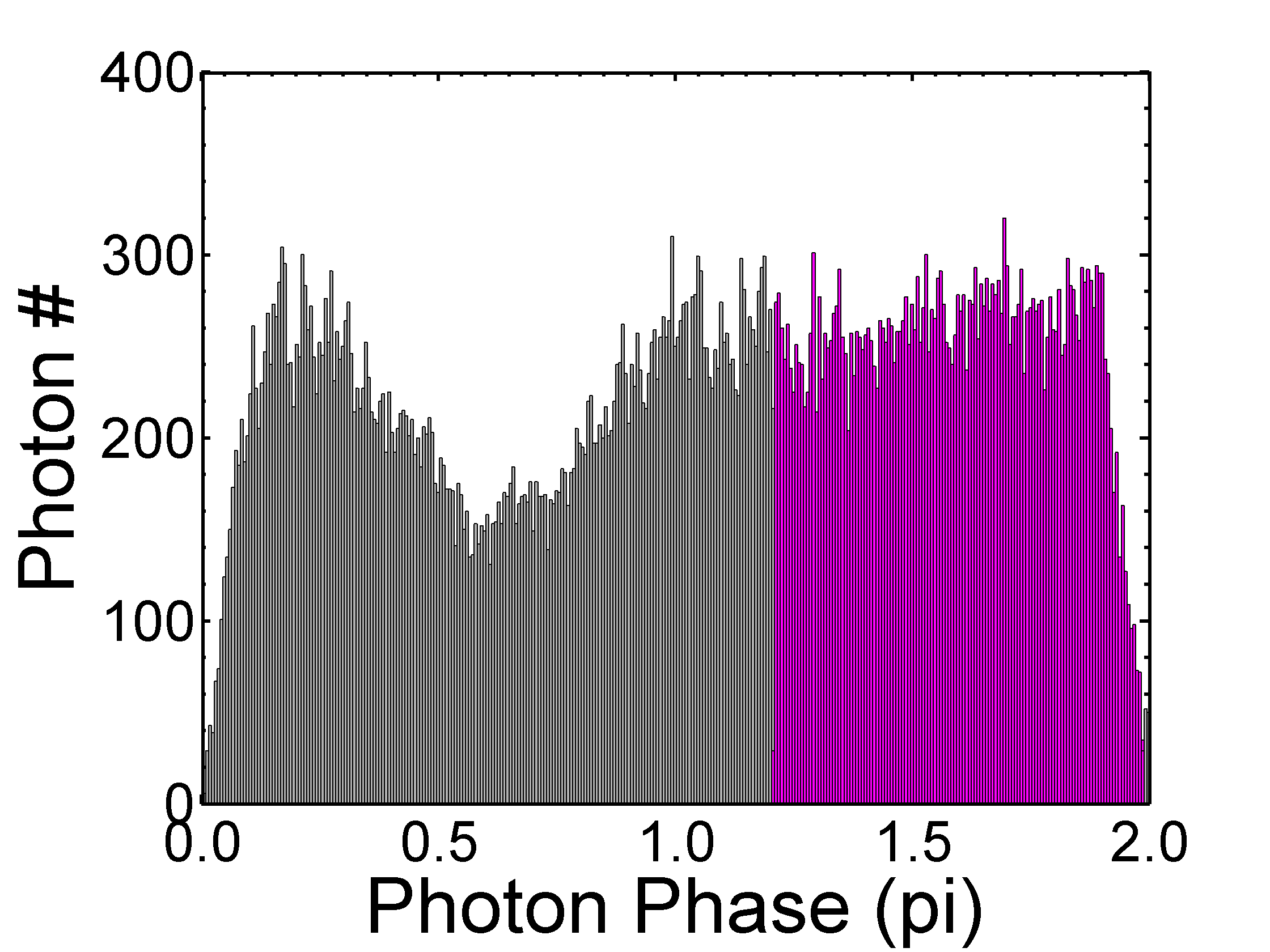}& \multicolumn{2}{|c}{} \\\cline{1-2}
\end{tabular}
\caption{The measured ellipsoid shape vs the
the scattered photons phase range. The figures are given
for a phase range of (a)40$^\circ$ (b)50$^\circ$ (c)60$^\circ$
(d)70$^\circ$ (e)80$^\circ$ (f)90$^\circ$ (g)100$^\circ$
(h)110$^\circ$ (i)120$^\circ$ (j)130$^\circ$ (k)140$^\circ$. Marked in purple are the
corresponding photon phase window.}\label{ZeurekEllipsoidChange}
\end{figure}
\end{center}

\begin{center}
\begin{figure}
\includegraphics[scale=0.5]{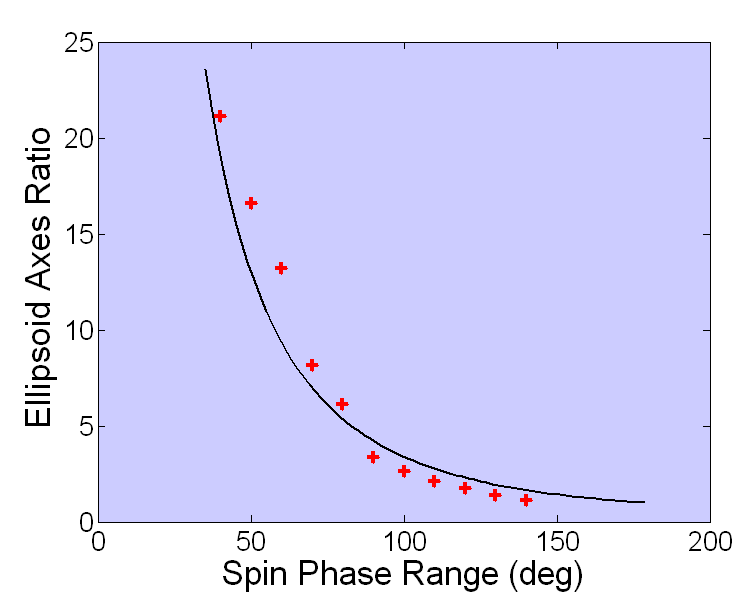}
\caption{The measured ellipsoid aspect ratio, in
the $\hat{x}-\hat{y}$ plane, vs the scattering
events phase range. Red crosses show measurements. The black line is a
numerical simulation result.} \label{RotatingZurek3}
\end{figure}
\end{center}
\end{widetext}
%\begin{thebibliography}{}
%\bibitem{Akerman2011} Akerman N, Kotler S, Glickman Y, Ozeri R (2011) Quantum correction of photon-scattering errors arXiv:1111.1622
%\bibitem{IkeAndMike} Nielsen MA, and Chuang (2000) IL Quantum Computation and Quantum Information, Campridge University Press
%\end{thebibliography}